\documentclass[%
 reprint,
 amsmath,amssymb,
 aps,
pre,
]{revtex4-1}

\usepackage{graphicx}
\usepackage{dcolumn}
\usepackage{bm}

\begin{document}


\title{Lattice Boltzmann model for the simulation of the wave equation in curvilinear coordinates.}

\author{A.M. Velasco}\email{amvelascos@unal.edu.co}
 \affiliation{Simulation of Physical Systems Group, Department of Physics, Universidad Nacional de Colombia, Cra 30 No. 45-03, Ed. 404, Of. 348, Bogotá D.C., Colombia}
  \affiliation{
 Computational Physics for Engineering Materials, Institut f\"ur Baustoffe, Eidgenössische Technische Hocschule (ETH) - Z\"urich, Wolfgang-Pauli-Str. 27, 8093 Z\"urich, Switzerland.}
\author{J. D. Mu\~noz}%
 \email{jdmunozc@unal.edu.co}
\affiliation{%
 Simulation of Physical Systems Group, Department of Physics, Universidad Nacional de Colombia, Cra 30 No. 45-03, Ed. 404, Of. 348, Bogotá D.C., Colombia}%
\author{M. Mendoza}
\affiliation{
 Computational Physics for Engineering Materials, Institut f\"ur Baustoffe, Eidgenössische Technische Hocschule (ETH) - Z\"urich, Wolfgang-Pauli-Str. 27, 8093 Z\"urich, Switzerland.
}%

\date{\today}

\begin{abstract}
Since its origins, lattice-Boltzmann methods have been restricted to rectangular coordinates, a fact which jeopardises the applications to problems with cylindrical or spherical symmetries and complicates the implementations with complex geometries. However, M. Mendoza \cite{MENDOZA} recently proposed in his doctoral thesis a general procedure (based on Christoffel symbols) to construct lattice-Boltzmann models on curvilinear coordinates, which has shown very good results for hydrodynamics on cylindrical and spherical coordinates. In this work, we construct a lattice-Boltzmann model for the propagation of scalar waves in curvilinear coordinates, and we use it to determine the vibrational modes inside cylinders, trumpets and tori. The model correctly reproduces the theoretical expectations for the vibrational modes, and exemplifies the wide range of future applications of lattice-Boltzmann models on general curvilinear coordinates.
\end{abstract}

\pacs{Valid PACS appear here}
\maketitle


\section{\label{Intro}Introduction}
Lattice-Boltzmann methods (LBM) were first introduced as mesoscopical models to simulate a wide variety of processes in fluid dynamics, from wind tunnels and turbulence to porous media and general rheology \cite{Succi2001}; but later on they were extended to more general systems, like waves  {\cite{Chopard1997,Chopard2009,Guangwu2000}, electrodynamics \cite{Mendoza2010}, or even Quantum Mechanics \cite{Palpacelli2008} and, therefore, they can be considered nowdays as a general numerical scheme to solve differential equations that can be written as a set of conservation laws. In comparison with other numerical schemes (like finite-differences or finite-element methods) all variables needed to compute for the next time step at a single node are present in the node itself (before being moved to the neighbouring cells), making them perfect to run parallel on graphic cards. Additionally, LBM models like the one for electrodynamics, has shown to be up to five times faster than the mentioned methods even in serial calculations for the same precision order \cite{Mendoza2010}. Because of these advantages, LBMs have gained the interest of a wide range of research areas and industrial applications; however, since most lattice-Boltzmann models assume a homogeneous and isotropic set of velocity vectors to move the information from node to node, the computational domain has been restricted to a rectangular array of cubic cells, forcing the use of staircase approximations on curved boundaries and imposing three-dimensional simulation domains for systems that, because of axial or spherical symmetries, were essentially two-dimensional, with an exorbitant increase in  computational costs. This is also the case of simulating waves with lattice Boltzmann on acoustic systems. Symphonic and traditional instruments like violins, trumpets or drums, modern auditoriums, complex geological wells in seismic prospection and hearing organs like the Cochlea have too complex geometries to be properly described in Cartesian coordinates, asking for the need of new LBM for acoustics in generalized coordinates to take advantage of their versatility and parallel nature.

The simulation of waves is a wide area of research by itself. Waves are present almost in every phenomenon, from the surface patterns on water and the propagation of sound in solids and fluids to electromagnetic potentials and gravitational waves, and not all are related with fluids. Moreover, even in the case of fluids, focusing directly on reproducing the wave equation of interest reduces the complexity and computational costs, in comparison to simulate the whole fluid mechanics problem. The first approach to simulate the wave equation by LBM was proposed in 1998 by Chopard, Luthi and Wagen \cite{Chopard1997}. The model modifies the equilibrium distribution function and the macroscopic variables to reproduce the wave equation, and with simple modifications runs on almost every velocity set, including D2Q5 and D3Q7. Later, in 2000 Guangwu \cite{Guangwu2000} proposed to redefine the first macroscopic momentum of the LBM as the temporal derivative of the wave pressure. This also leads to the pressure wave equation, but an additional integration step is needed. In addition, some approaches use the standard LBM to simulate the Navier-Stokes equations (NSE) and meassure the pressure waves present in such dynamics, but with some restrictions and at a very high computational cost. In 2011, Li and Shan \cite{Li2011} simulated the NSE  with a Multiple Relaxation Time lattice-Boltzmann and studied the decaying of pressure signals due to the viscous, thermal and acoustic damping. Their model requires very high order approximations, which make it very demanding on computational resources, and applies only for low frequencies. In 2015 Sun \textit{et. al.} \cite{Sun2015} used a forced lattice-Boltzmann model to solve the NSE for the fluid inside a fully saturated porous medium, and studied the propagation of a pressure front there. The absorption coefficients derived from the simulations coincide well with the theoretical expectations, but not with the experiments, due, perhaps, to the need of extra factors like thermal dissipation and flexibility of the porous medium. In 2016, Salomons, Lohman and Zhou \cite{Salomons2016}, studied various cases of outdoor acoustics in 2D with a Multiple Relaxation Time lattice-Boltzmann model for the NSE, but the restrictions of minimal viscosity in the model avoids them to obtain the almost non-dissipative behaviour found in the real air. These limitations in the use of using LMBs for the NSE to simulate acoustics favours the use of simpler schemes directly designed for the wave equation, as the one proposed by Chopard \textit{et. al}.

Various approaches have been proposed to overcome the Cartesian restriction in LBMs. One of the first proposals was made by Nannelli and Succi \cite{Nannelli1992}, consisting in a finite volume solution of the kinetic Boltzmann equation where the two-dimensional cells can be irregular quadrilaterals; but this scheme introduces a new coarse-grained distribution function (averaging the standard lattice-Boltzmann distributions) that should be retrieved through additional interpolation steps. Later, in 1997, He and Doolen \cite{He1997} implemented a lattice-Boltzmann scheme in polar coordinates, based on an algorithm proposed by He, Luo and Dembo \cite{He1996}, where that additional interpolation step is still required. In 1998 a grid refinement scheme with boundary-fittings for complicated geometries was proposed by Filippova and H{\"a}nel  \cite{Filippova1998}. In this work, smaller cells are located in regions where a higher resolution is needed. Even though this strategy reduces some computational costs, the staircase approximation does not disappear and some interpolations on the curved boundaries of complex systems are still needed. Up to now, the implementation of lattice-Boltzmann models in curved general geometries has been a hard task, not only because of the insertion of additional computing steps, but mainly because each new geometry needs a new cautious discretization scheme to ensure a correct definition of the boundary conditions. In 2010 and 2015 Li \textit{et. al.}\cite{Li2010} and Reijers, Gelderblom and Toschi,\cite{Reijers2016}, respectively, developed a lattice-Boltzmann scheme for fluid mechanics to be implemented on axisymmetric coordinate systems. In particular, Reijers and co-workers constructed a bi-phasic lattice-Boltzmann model capable to achieve density ratios between phases up to 1000 and used it to study the propagation of pressure waves in such coordinates, but with an excessive dissipation due to numerical viscosity. In 2012 a more general approach in two dimensions was developed by Budinsky  for both the shallow water and the Navier-Stokes equations\cite{Budinski2012}. In that scheme, the equations are written in general coordinates, and the additional geometric terms (containing the Jacobian and Jacobian spatial derivatives) are introduced as forcing terms in the collision operator. The main advantage of this model is that all information about the curvilinear coordinates is included in the equilibrium distribution function and the forcing term, and no further discretization or interpolation steps are needed.
Simultaneously with the Budinsky proposal, M. Mendoza \cite{MENDOZA}  introduced in his doctoral thesis a new strategy to built lattice-Boltzmann models for fluids on any three-dimensional curvilinear coordinate system. The strategy also reproduces in the macroscopic limit the desired equations in generalized coordinates, but using the metric tensor and Christoffel symbols instead of the Jacobian . In contrast with Budinsky's method, the forcing terms are included both in the equilibrium functions and in the macroscopic quantities, following the procedure by Guo \textit{et. al.} \cite{Guo2002} and reaching second-order accuracy. Again, the strategy does not require neither a specific discretization scheme for each problem nor additional interpolation steps. This model has been successfully used to study the Dean's instability in ellipsoidal coordinates \cite{Debus2014}, the flow through randomly curved media \cite{Mendoza2013} and, more recently, the energy dissipation due to curvature \cite{Debus2017}.

In this work we extend a modified version of the model by Chopard \textit{et. al} \cite{Chopard1997,Chopard2009} to simulate acoustic waves on generalized coordinates by following the proposal of M. Mendoza \cite{MENDOZA}. The method, that can be used on any coordinate system, was tested by simulating the normal modes inside a cylinder, a trumpet and a torus, finding second-order accuracy. Section \ref{Cartesian} reviews the LBM proposed by Chopard \textit{et. al} and derives an alternative form by using Hermite polynomials, comparing their performance in the simple case of a point source in two dimensions. Section \ref{Generalized} extends that alternative form to generalized coordinates, including the general recipe to build the LBM for waves on any coordinate system. The model is tested in Section \ref{Systems} by simulating the acoustic waves inside a cylinder, a trumpet and a torus. Finally, Section \ref{Conclusions} summarizes the main results and conclusions. Videos of the simulations can be found in the supplementary material attached to this manuscript.
\section{LBM for waves in Cartesian coordinates}
\label{Cartesian}
Let us start from the lattice-Boltzmann's equation with the Bhratnagar-Gross-Krook approximation \cite{Bhatnagar1954}, 
\begin{equation}
\begin{split}
f_i\left(\vec{x}+\vec{\xi}_i\delta_{t},\,\vec{\xi}_i,\, t+\delta_{t}\right)-f_i\left(\vec{x},\,\vec{\xi}_i,\, t\right)=\\
-\frac{\delta_{t}}{\tau}\left[f_i\left(\vec{x},\,\vec{\xi}_i,\, t\right)-f_i^{eq}\left(\vec{x},\,\vec{\xi}_i,\, t\right)\right]\, ,\label{eq:evolution}
\end{split}
\end{equation}
where $\delta_t$ is the time step, we choose $\delta_t=1$ hereafter, $\tau$ is a characteristic relaxation time and $f_i\left(\vec{x},\,\vec{\xi}_i,\, t\right)$ is the distribution function assigned to the velocity vector $\vec{\xi}_i$ \footnote{In lattice-Boltzmann models for fluid mechanics, this distribution function is proportional to the probability to find a molecule at position $\vec{r}$ and time $t$ with velocity $\vec{\xi}_i$, but it is just a system's variable in the general case}. If the new distribution function (i.e. the first term on the left)  were written at the same place, instead of moving to the neighbouring cells, this equation would represent an exponential decay to equilibrium. The left-hand side represents the time evolution of the distribution function without external forces and is called the {\it advection} term, whereas the right-hand side represents the interactions among distribution functions and is called the {\it collision} term. Note that the velocity space is reduced to a discrete set of velocity vectors $\vec\xi_i$, usually with some associated weights $w_i$, like the set D3Q7 described in  Fig. \ref{Fig.VEL}. 
The macroscopic quantities are computed in terms of the discrete velocities and the distribution functions, 
\begin{equation}\label{MacroscopicQuantities}
	P=\sum_i f_i \quad , \quad
	\vec J=\sum_i f_i \vec \xi_i \quad .
\end{equation}
In the case of acoustic waves, $P$ represents the pressure and $\vec J=-B\frac{\partial D}{\partial t}$ is proportional to the time derivative of the mean particle displacement $D$, with $B$ the bulk modulus.

\begin{figure}
\centering
 \includegraphics[width=0.4\textwidth]{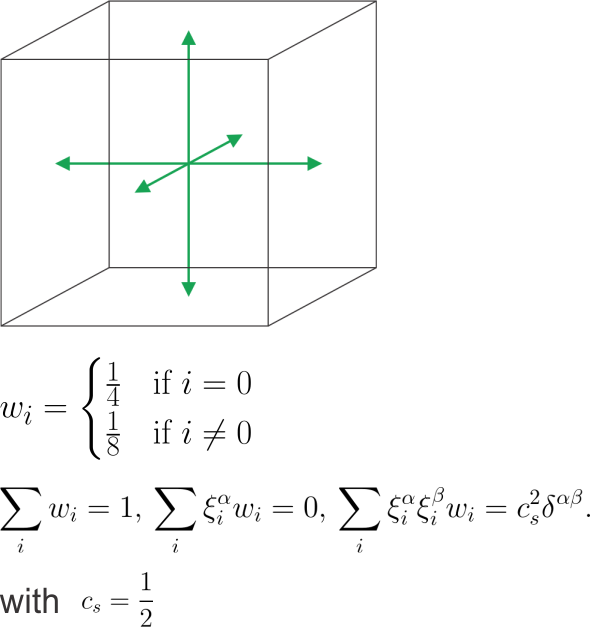}
\caption{Discrete set of 7 velocities in 3 dimensions, its weights and isotropy relations}
\label{Fig.VEL}
\end{figure}

The set of conservation laws the system fulfils in the macroscopic limit can be found by performing a Chapman-Enskog expansion \cite{Chopard2009}, of the evolution law Eq. (\ref{eq:evolution}). The left-hand side of such equation is replaced by a Taylor expansion up to second order and, both the distribution functions $f_i=f_i^{(0)}+\epsilon f_i^{(1)}+\epsilon^2 f_i^{(2)}$ and the differential operators $\nabla=\epsilon\nabla_1$ and $\frac{\partial}{\partial t}=\epsilon\frac{\partial}{\partial t_1} +\epsilon^2\frac{\partial}{\partial t_2}$ are expanded on the Knudsen's number $\epsilon$, such that $\epsilon\to 0$ is the continuous limit. By replacing and equating order by order, one obtains  
\begin{equation}
f_i^{(0)}=f_i^{(eq)}\quad,\label{ZeroOrder}
\end{equation}
\begin{equation}
-\frac{1}{\tau}f_i^{(1)}=\left[\frac{\partial}{\partial t_1} +\vec \xi_i\cdot\vec\nabla_1\right] f_i^{(0)}\quad,
\label{FirstOrder}
\end{equation}

\begin{equation}
\frac{\partial}{\partial t_2}f_i^{(0)}+\left(1-\frac{1}{2\tau}\right)\left[\frac{\partial}{\partial t_1} +\vec \xi_i\cdot\vec\nabla_1\right] f_i^{(1)}=\frac{1}{\tau}f_i^{(2)}
\label{SecondOrderComplete}
\end{equation}
where we have used Eq. \ref{FirstOrder} to rewrite Eq. \ref{SecondOrderComplete}. If one chooses $\tau=1/2$, Eq \eqref{SecondOrderComplete} reduces to.
\begin{equation}
-\frac{1}{\tau}f_i^{(2)}=\frac{\partial}{\partial t_2} f_i^{(0)}\quad.
\label{SecondOrder}
\end{equation}
By multiplying Eq. \ref{FirstOrder} by $\epsilon$ and Eq. \ref{SecondOrder} by $\epsilon^2$ and adding both equations, one recovers the time and space derivatives, obtaining
\begin{equation}
	-\frac{1}{\tau}\left(\epsilon f_i^{(1)}+\epsilon^2 f_i^{(2)}\right)=\frac{\partial}{\partial t} f_i^{(0)}
	+\vec\nabla\cdot\left(\vec \xi_i f_i^{(0)}\right)\quad ,
\label{ConservationLaws}
\end{equation}
where we use the fact that the velocity vectors $\vec\xi_i$ are the same from cell to cell. Now, by summing over the whole set of discrete velocities $\vec{\xi}_i$ and taking the limit $\epsilon\to 0$, we find a first conservation law
\begin{equation}\label{1Conservation}
	\frac{\partial P}{\partial t}+\nabla\cdot\vec J=0\quad \quad .
\end{equation}
Similarly, multiplying by $\vec{\xi}_i$ before summing give us a second conservation law
\begin{equation}\label{2Conservation}
	\frac{\partial\vec J}{\partial t}+\nabla\cdot\vec \Pi^{(0)}=0\quad,
\end{equation}
where we defined $\Pi^{(0)}=\sum_i f_i^{eq} \vec \xi_i \otimes \vec \xi_i$.

Our goal now is to define the equilibrium distribution functions in such a way that those two conservation laws combine to obtain the scalar wave equation. Note that if we handle to obtain 
\begin{equation}
{\Pi^{(0)}}^{\alpha\beta}=\sum_if_i^{eq}\xi_i^\alpha \xi_i^\beta=c^2P\delta^{\alpha\beta}\quad,
\label{Stress}
\end{equation}
Eq. (\ref{2Conservation}) becomes
\begin{equation}\label{2ConservationReplaced}
	\frac{\partial\vec J}{\partial t}+c^2\nabla P=0\quad,
\end{equation}
Now, by taking the time derivative of Eq. \eqref{1Conservation} and replacing into Eq. \eqref{2ConservationReplaced} one obtains
\begin{equation}\label{waveEuclidean}
	\frac{\partial^2 P}{\partial t^2}-c^2\nabla^2 P=0\quad,
\end{equation}
that is, the wave equation for the pressure $P$ in Cartesian coordinates.

Finally, it is possible find the proper functional form for the equilibrium distribution in two ways: The first one consists in proposing an Ansatz for the equilibrium distribution to obtain the desired form for ${\Pi^{(0)}}^{\alpha\beta}$ and complete it to reproduce the macroscopic quantities Eq.\eqref{MacroscopicQuantities} with $f_i^{eq}$ instead of $f_i^{0}$. If the velocity vectors $\vec\xi_i$ and weights $w_i$ fulfil $\sum_iw_i\xi_i^\alpha \xi_i^\beta=c_s\delta^{\alpha\beta}$, one can propose
\begin{equation}\label{feqMomentMatchingCartesian}
f_i^{eq}=\begin{cases}
P+\frac{c^2}{c_s^2}P\left(w_0-1\right)&{\text if\quad} i=0\\
\frac{w_i}{c_s^2}\left(c^2P+\vec{\xi}_i\cdot\vec{J}\right)&{\text if\quad} i\neq0\
\end{cases}\, .
\end{equation}
where the additional terms are added to obtain $P=\sum_i f_i^{eq}$ and $\vec J=\sum_i f_i^{eq} \vec \xi_i$. This is the equilibrium function proposed by Chopard \textit{et. al.} \cite{Chopard1997}.
The second one consists in discretizing a proposed continuous equilibrium distribution function as a truncated a Hermite polynomial series \cite{Kruger2017}. Let propose the continuous form
\begin{equation}\label{continuousfeq}
f^{eq}\left(\vec{\xi}\right)=\frac{P}{(\sqrt{2\pi c_s^2})^D}\text{e}^{-(\vec{J}-\vec{\xi})^2/2c_s^2}\, .
\end{equation}
which lead us to the discrete equilibrium distribution function (see Appendix\ref{AppHermite}) 
\begin{equation}\label{EquilibriumHermite}
f_i^{eq}=\begin{cases}
P-\left(\frac{5P}{2}-\frac{3c^2P}{2c_s^2}\right)\left(1-w_0\right)+\\
\frac{3c^2P}{2c_s^2}\left(c^2-c_s^2\right)&\text{if } i=0\\
\\
w_{i}\bigg[P+\frac{\xi_i\cdot J}{c_s^2}\\
+\frac{P}{2c_s^4}\left(c^2-c_s^2\right)\left(\xi_i^2-3c_s^2\right)\bigg] & \mbox{otherwise}
\end{cases}\quad,
\end{equation}
where the $f_0^{eq}$ contribution was found from $f_0^{eq}=P-\sum_{i=1}^Q f_i^{eq}$ (hereafter, Einstein summation is assumed).
 
Fig. \ref{Fig.ComparisonCartesian} shows the wave patterns obtained through both methods for a point source in two dimensions. The simulation domain was $100\times 100\times 1$ cells, the wave speed was set to $c=0.5$ cells per timestep and the source at the central cell oscillates as $P=A\sin(\omega t)$, with $\omega=2\pi/\lambda$ and $\lambda=10$ cells. One can observe that both approaches reproduce the theoretical profile $P(r)=A\mathcal{J}_0\left(kr\right)\sin\left(\omega t\right)$, with RMS errors of 0.0487 for the Moment Matching approach and 0.0484 for the one obtained from Hermite polynomials expansion. Although the first approach is simpler, is the second the one we will use as starting point for our LBM in generalized coordinates.
\begin{figure}
\centering
 \includegraphics[width=0.5\textwidth]{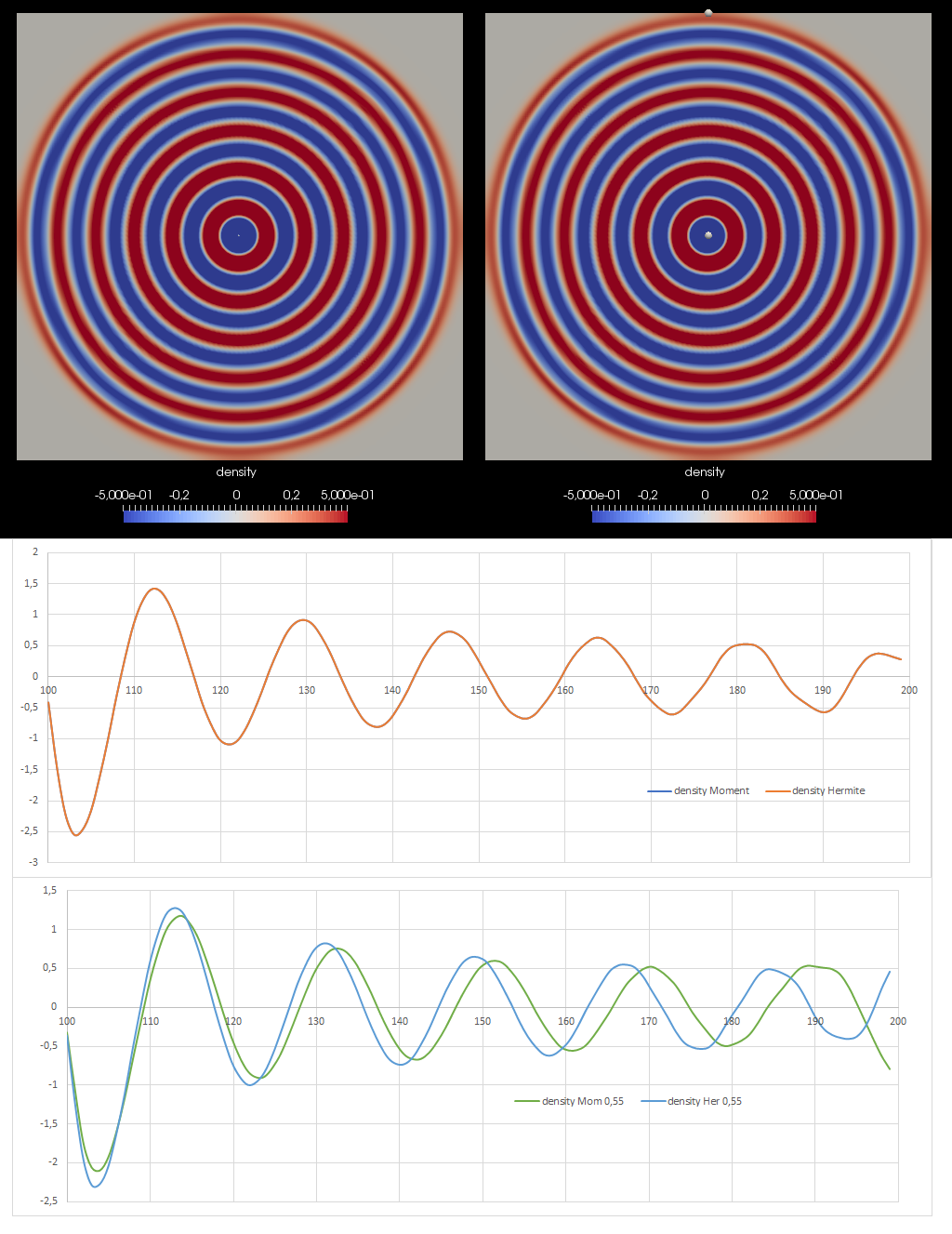}
\caption{Pressure profiles of a circular wave simulated by using the Eqs. \ref{feqMomentMatchingCartesian} and \eqref{EquilibriumHermite} respectively (top), radial amplitude at a given time obtained from both models compared with the theoretical expectation (middle) and the comparison but at a higher wave velocity.} 
\label{Fig.ComparisonCartesian}
\end{figure}
   
\section{\label{Generalized}LBM for waves in generalized coordinates}

Our strategy to build a LBM in generalized coordinates is to maintain a Cartesian array of cells to store the information in the computer, each dimension in this array representing a generalized coordinate in the real space. So, the velocity vectors and weights will be the same as in a traditional cubic discretization scheme. In contrast, the  macroscopic quantities, equilibrium functions and forcing terms are chosen to simulate the conservation laws in generalized coordinates, as they were just novel and complicated equations to be represented on that Cartesian scheme. 

Let us start by writing the differential operators in its generalized form and considering the additional terms as forcing terms. The wave equation in general coordinates reads
\begin{equation}
\frac{\partial^2P}{\partial t^2}-\frac{c^2}{\sqrt{g}}\partial_i\left(\sqrt{g}g^{ik}\partial_k P\right)=0\quad,
\label{waves}
\end{equation}
where $g^{kl}$ is the contravariant metric tensor of the geometry and $g$, its determinant. Here, the second term is the explicit definition of the Laplace-Beltrami operator applied on the scalar quantity $P$. The gradient of a scalar function and the divergence of vector and tensor fields are given by \cite{Lawden2002}
\begin{equation}
(\nabla P)^l =\partial^l P= g^{lk}\partial_k P\quad,
\end{equation}
\begin{equation}
\nabla \cdot \vec{J} =\frac{1}{\sqrt{g}}\partial_i(\sqrt{g}J^i)\quad,
\end{equation}
\begin{equation}
\nabla_j \left(\Pi^{(0)}\right)^{ij}= \frac{1}{\sqrt{g}}\partial_j\left(\sqrt{g}\left(\Pi^{(0)}\right)^{ij}\right)+\Gamma^i_{jk}\left(\Pi^{(0)}\right)^{jk}\quad.
\end{equation}
Thus, the conservation laws (Eqs. \eqref{1Conservation} and \eqref{2Conservation}) now read 
\begin{equation}\label{1Conservationcurved}
	\frac{\partial\left(\sqrt{g} P\right)}{\partial t}+\partial_i(\sqrt{g}J^i)=0\quad \quad ,
\end{equation}
\begin{equation}\label{2ConservationcurvedForce}
	\frac{\partial \left(\sqrt{g} J^i\right)}{\partial t}+\partial_j\left(\sqrt{g}\left(\Pi^{(0)}\right)^{ij}\right)=-\Gamma^i_{jk}\left(\Pi^{(0)}\right)^{jk} \sqrt{g}\quad.
\end{equation}
Note that the right-hand side in Eq. \eqref{2ConservationcurvedForce} can be considered as a source term of the conservation law, it is, therefore, a forcing term in the lattice-Boltzmann scheme. The forcing term can be treated as usual by following the approach by Guo \textit{et. al.} \cite{Guo2002} with $\tau=1/2$ as shown below (Appendix \ref{ChapmannEnskogAppendix}). 
Observe that, if we handle to find an equilibrium distribution function such that the stress tensor were
\begin{equation}
\left(\Pi^{(0)}\right)^{ij}=c^2Pg^{ij}\, ,
\label{StressCurved}
\end{equation}
Eq.(\ref{2ConservationcurvedForce}) would be
\begin{equation}\label{Gradient}
	\frac{\partial \left(\sqrt{g} J^i\right)}{\partial t}=-\partial_j\left(\sqrt{g}c^2Pg^{ij}\right)-\Gamma^i_{jk}c^2Pg^{jk} \sqrt{g}\quad
\end{equation}
and, by taking the time derivative of Eq. (\ref{1Conservationcurved}),
\begin{equation}\label{ConservationLaw}
	\frac{\partial^2\left(\sqrt{g} P\right)}{\partial t^2}+\partial_i\left(\frac{\partial \left(\sqrt{g} J^i\right)}{\partial t}\right)=0\quad \quad ,
\end{equation}
the left hand side of Eq. (\ref{Gradient}) could be replaced in the second term of Eq. (\ref{ConservationLaw}) to obtain
\begin{equation}\label{ConservationLaw2}
	\frac{\partial^2\left(\sqrt{g} P\right)}{\partial t^2}-c^2\partial_i\bigg(\partial_j\left(\sqrt{g}Pg^{ij}\right)+\Gamma^i_{jk}g^{jk} \sqrt{g}P\bigg)=0\quad \quad .
\end{equation}
This is exactly the wave equation in general coordinates (Eq. \ref{waves})  we want to reproduce.

Next, because the macroscopic fields $P$, $\vec J$ and $\Pi^{(0)}$ in  Eq. (\ref{ConservationLaw}) and (\ref{ConservationLaw2}) appear multiplied by $\sqrt{g}$, let us define the statistical moments of the distribution functions such that 
\begin{equation}\label{PressureCurvedSystem}
\sqrt{g}P=\sum_l f_l^{eq} \quad ,
\end{equation}
\begin{equation}\label{MomentumCurvedSystem}
\sqrt{g}J'^i=\sum_l f_l^{eq}\xi^i_l - \frac{1}{2}c^2P\Gamma^i_{jk}g^{jk}\quad,
\end{equation}
\begin{equation}\label{StressCurvedSystem}
\sqrt{g}\left(\Pi^{(0)}\right)^{ij}=\sum_l f_l^{eq}\xi^i_l\xi^j_l=\sqrt{g}c^2Pg^{ij}\, .
\end{equation}
The additional term in the righ-hand side of the macroscopic quantity $\vec{J}'$ comes from the forcing term of the conservation law (Eq. (\ref{2ConservationcurvedForce})), as described by Guo \textit{et. al.} for $\tau=1/2$.

The problem reduces again to find the equilibrium distribution functions such that Eq. (\ref{PressureCurvedSystem}-\ref{StressCurvedSystem}) hold. First, we write a continuous equation that is just the extension of Eq. \eqref{continuousfeq} 
\begin{equation}\label{continuousfeqGeneral}
f^{eq}\left(\vec{\xi}\right)=\frac{P}{(\sqrt{2\pi c_s^2})^D}\text{e}^{-[g_{\alpha\beta}\left(J^\alpha-\xi^\alpha\right)\left(J^\beta-\xi^\beta\right)]/2c_s^2}\, .
\end{equation}
Up to second order, the obtained form of the discrete equilibrium distribution function for the wave equation in curvilinear coordinates is (see Appendix \ref{AppHermite})
\begin{equation}\label{EquilibriumCurvedSystemHermite}
f_i^{eq}=\begin{cases}
w_0\sqrt{g}P\left[\frac{1}{w_0}-\frac{c^2}{2c_s^2}g^{\alpha\alpha}\right]-\sqrt{g}P+\frac{5}{2}\sqrt{g}Pw_0&\text{if } i=0\\
\\
w_i\sqrt{g}\bigg[P+\frac{\vec{\xi}_i\cdot\vec{J'}}{c_s^2}\\
+\frac{P}{2c_s^4}\left(g^{\alpha\beta}c^2-c_s^2\delta^{\alpha\beta}\right)\left(\xi_i^\alpha\xi_i^\beta-c_s^2\delta^{\alpha\beta}\right)\bigg]& \text{if } i\neq 0
\end{cases}\, .
\end{equation}
This form of the equilibrium distribution function evidences the existence of a conflictive term, the factor $\left(g^{\alpha\beta}c^2-c_s^2\delta^{\alpha\beta}\right)$, which accounts for how different is the metric of the involved curvilinear coordinate system from the Cartesian one. If the difference is too high, this term will produce  numerical instabilities, inducing negative values for the equilibrium distributions. 
To diminish that effect, we move that term from the equilibrium function to the forcing term in $\vec J$, which becomes
\begin{equation}\label{JCorrectionForce}
\begin{split}
\sqrt{g}J'^i=&\sum_l f_l^{eq}\xi^i_l \\
&-\frac{1}{2}c^2P\Gamma^i_{jk}g^{jk}+\frac{1}{2}\partial_j\left[\sqrt{g}c^2P\left(g^{ij}-\delta^{ij}\right)\right]\quad,
\end{split}
\end{equation}
the whole forcing term is, therefore
\begin{equation}
\mathcal{F}^i=- \frac{1}{2}c^2P\Gamma^i_{jk}g^{jk}+\frac{1}{2}\partial_j\left[\sqrt{g}c^2P\left(g^{ij}-\delta^{ij}\right)\right]\quad,
\end{equation}
and the equilibrium function reduces to
\begin{equation}\label{EquilibriumCurvedSystemHermiteCorrected}
f^{eq}=\begin{cases}
w_{0}\sqrt{g}P & \mbox{for\,}i=0\\
w_{i}\sqrt{g}\left[P+\frac{\xi_i^kJ'^k}{c_s^2}\right] & \mbox{otherwise}
\end{cases}\quad.
\end{equation}
The spatial derivative $\partial_j\left[\sqrt{g}c^2P\left(g^{ij}-\delta^{ij}\right)\right]$ in Eq \eqref{JCorrectionForce} is easily computed as (See Appendix \ref{AppendixA})
\begin{equation}
\partial_j A^{ij}=\frac{1}{c_s^2}\sum_lw_l\xi_l^j A^{ij}\left(x^j+\xi_i^j\right)+\mathcal{O}\left(\delta x^2\right)
\label{DerivativeFromTaylor}
\end{equation}
where $A^{ij}=\sqrt{g}c^2P\left(g^{ij}-\delta^{ij}\right)$. Eq. \eqref{DerivativeFromTaylor} is obtained from a Taylor expansion in the discrete velocity space and, therefore, it has the same properties of the overall lattice-Boltzmann scheme \cite{Thampi2012}. 

Summarizing, the model consists in solving the Eq. \eqref{eq:evolution} with $\tau=1/2$ using the equilibrium distribution given in Eq. \eqref{EquilibriumCurvedSystemHermiteCorrected}, which in turn is computed from the updated macroscopic variables obtained from Eqs. \eqref{PressureCurvedSystem}, \eqref{JCorrectionForce} and \eqref{StressCurvedSystem}. Note that all the information about the curvature of the system is included in the product $\sqrt{g}$ present in the Eq. \eqref{EquilibriumCurvedSystemHermiteCorrected} and Eqs. \eqref{PressureCurvedSystem} and \eqref{StressCurvedSystem}, and in the forcing term of the Eq. \eqref{JCorrectionForce}, but the set of velocities $\{ \vec \xi_i \}$ remains the same. So, it is possible to adapt the geometry just by changing the metric tensor and Christoffel symbols, without altering the discretization scheme in the computer.

\section{Three examples: Cylinder, Trumpet and Torus}\label{Systems}

As examples, we will measure the normal modes for three cases: A cylinder, a trumpet and a torus. 
\subsection{Cylinder}
For the first case, cylindrical coordinates are given by the transformation 
\begin{equation}
\begin{split}
x&=r\cos\theta\\
y&=r\sin\theta\\
z&=z\\
\end{split}\quad,
\label{cylindricalsystem}
\end{equation}
as shown in Fig.  \ref{CCylinder}.
\begin{figure}
\centering
 \includegraphics[width=0.45\textwidth]{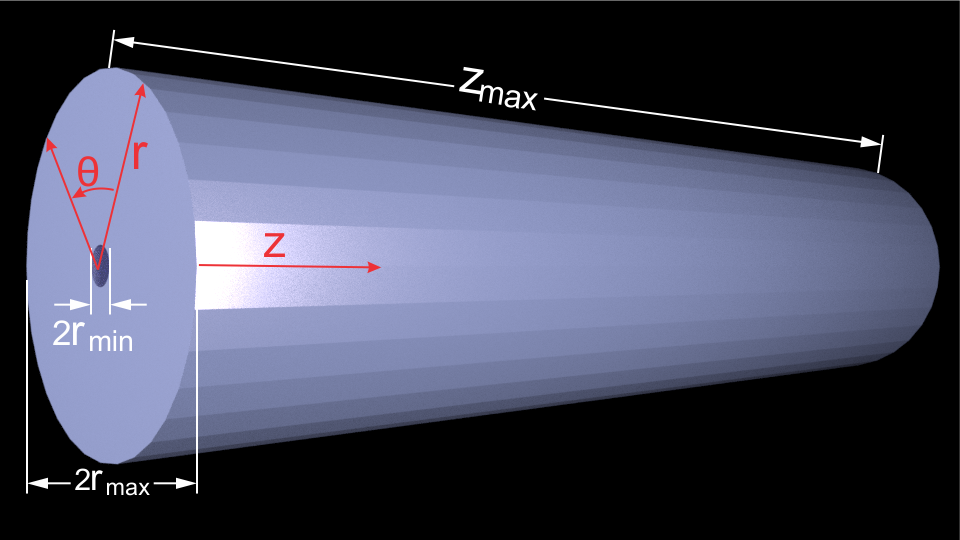}
\caption{Cylindrical coordinates representation.}
\label{CCylinder}
\end{figure}
The metric tensor reads:
\begin{equation}
g=\left(\begin{matrix}
1 & 0 & 0\\
0 & r^2 & 0\\
0 & 0 & 1\\
\end{matrix}\right)\quad,
\label{MetricCylindrical}
\end{equation}
and the non-zero Christoffel symbols are
\begin{equation}
 \Gamma^r_{\theta \theta}=-r,\quad  \Gamma^\theta_{r \theta}=\Gamma^\theta_{\theta r}=\frac{1}{r}\quad.
 \label{CylindricalChr}
\end{equation}
It is important to keep in mind that the information inside the computer is still stored in a cubic lattice; therefore, the visualization of the actual simulated system requires an inverse mapping (Fig.~\ref{Fig.Mapping}). 
\begin{figure}
\centering
 \includegraphics[width=0.3\textwidth]{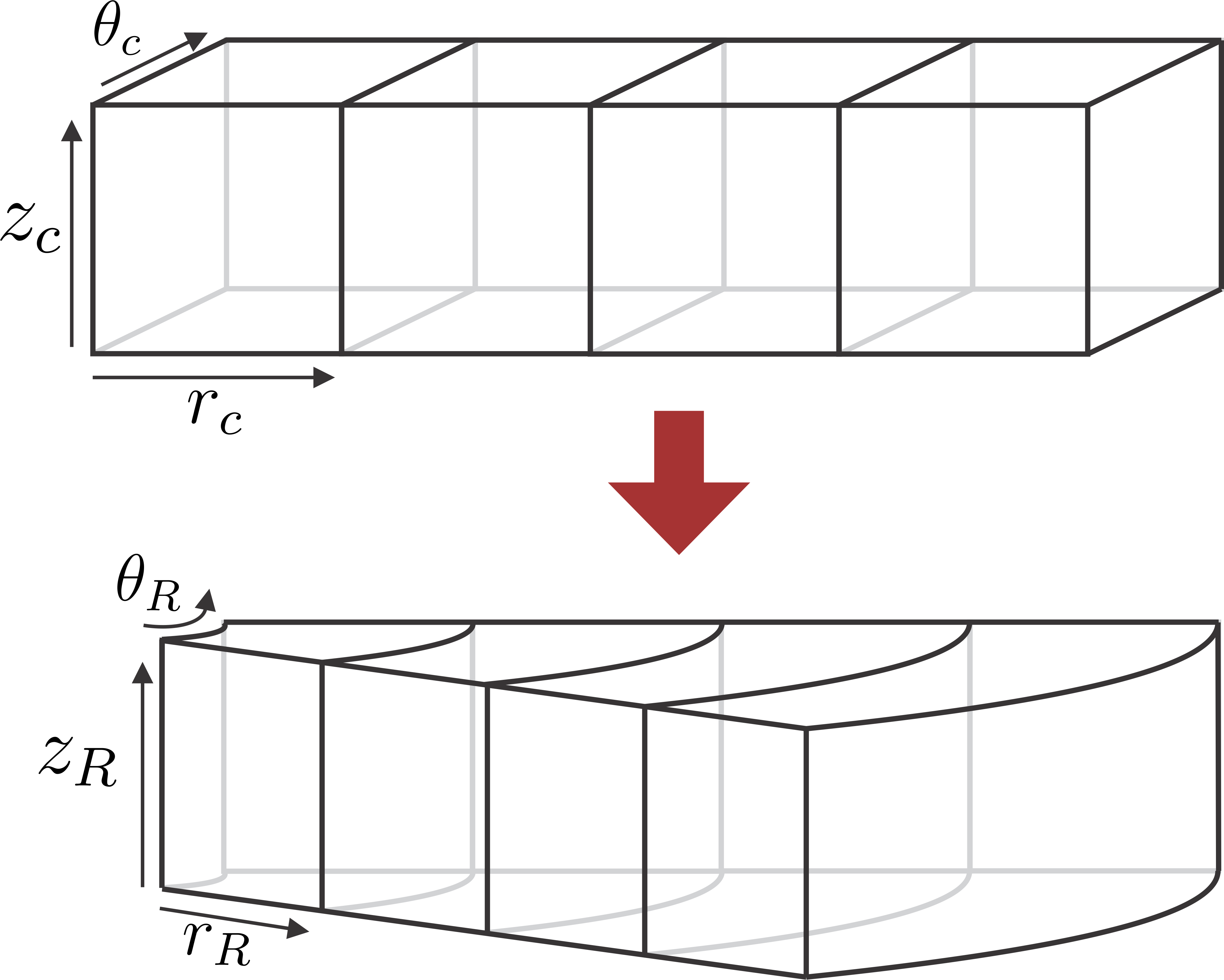}
\caption{Coordinate redefinition and transformation for a cylindrical system. The information is still stored on the computer in a cubic lattice. The subscript $C$ ($R$) indicates computer (real) coordinates.}
\label{Fig.Mapping}
\end{figure}
To validate the method, let us start by simulating a radial cylindrical wave expanding from the $z$ axis, whose theoretical solution is given by
\begin{equation}
\label{TheoreticalExpresion}
P(r,\theta,z)=A\mathcal{J}_0(R_{max}r) \approx \frac{A}{\sqrt{r}}\cos(r\pm ct)\quad ,
\end{equation}
where $\mathcal{J}_0$ is the 0-th order Bessel function. To perturbate the system, we impose the macroscopic quantity $P$ at the axis to be a harmonic perturbation,
 $
 P(r,\theta,z,t)_{source}=C\sin\left(\omega t\right)\delta(r-r_{\rm min}),
$
where $\delta$ is the Dirac's delta and $r_{\rm min}$ is the minimum value of $r$ to be simulated (since cylindrical coordinates have an indetermination at $r=0$).
For this simulation we use a cubic storage array of $100\times 3\times 10$ cells for the coordinates $r$, $\theta$ and $z$, respectively, with periodic boundary conditions in $\theta$ and $z$. For $r=r_{max}$ we imposed open boundary conditions (in other words, the simulation time is not enough for the wave to reach such boundary). Note that because of symmetry our field does not depend on $\theta$ and $z$, and the simulation could be one-dimensional in $r$ with a mesh  $L_r \times 1 \times 1$; but we choose to have some extra cells in $\theta$ and $z$ directions just to be sure that everything will run well for future simulations. The results for this benchmark (Fig. \ref{Fig.Radial}) are in excellent agreement with the theoretical predictions Eq. (\ref{TheoreticalExpresion}). In contrast with previous schemes \cite{Nannelli1992,He1996,Filippova1998}, no additional interpolation steps in the evolution are required. All the values of the quantities shown here and henceforth are in lattice units. for this simulation we have chosen $c=0.6$, $\omega=0.075$, $r_{\rm min}=1$ and $r_{\rm max}=500$.
\begin{figure}
\centering
\includegraphics[width=0.4\textwidth]{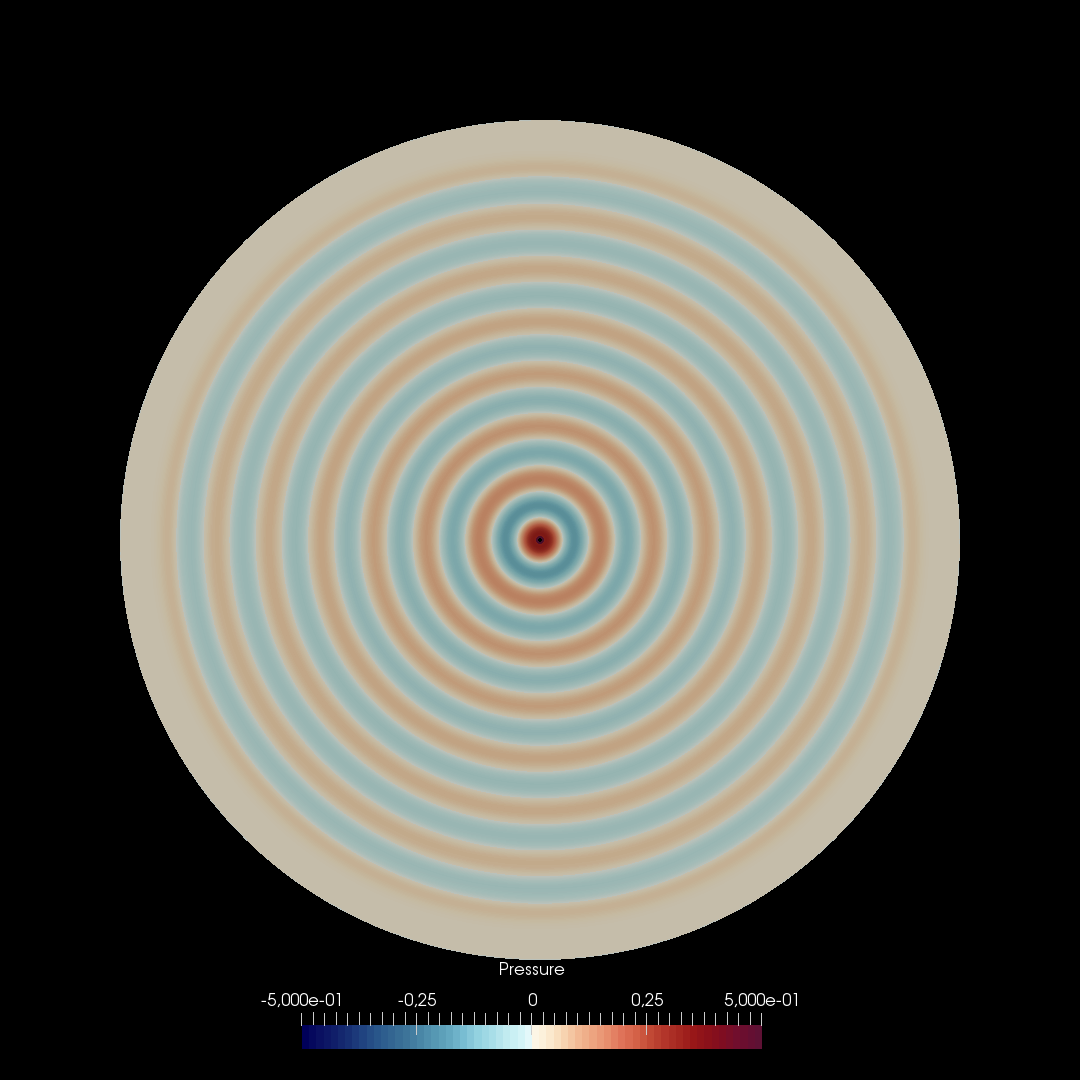}	 
 \includegraphics[width=0.45\textwidth]{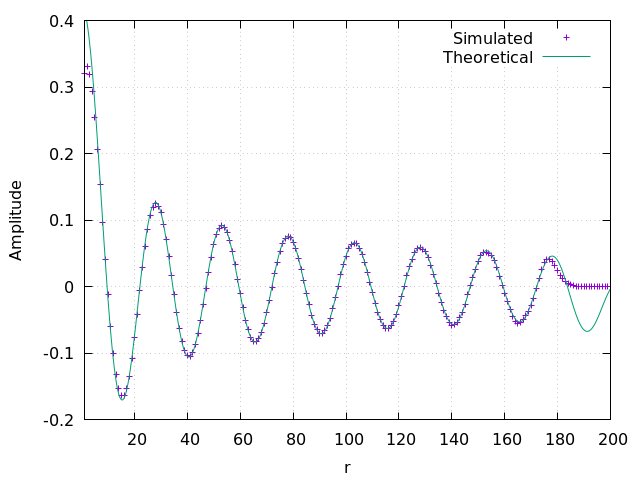} 
\caption{Simulation of a cylindrical wave in cylindrical coordinates. (Left) Pressure profile at a cross section of the cylinder at $t=365$ steps. (Right) Pressure as a function of the radial coordinate $r$ from the same simulation (purple dots) compared with the theoretical expression Eq.\eqref{TheoreticalExpresion} (green continuous line).}
\label{Fig.Radial}
\end{figure} 

Now, let us simulate an acoustic wave inside an open-closed end pipe whose dimensions in the computational domain in directions $r,\, \theta,\,z$ are $20 \times 5 \times 100$ cells with $r_{\rm max}=25$ and $z_{\rm max}=100$. Fig. \ref{Fig.BoundariesPipe} describes the boundary conditions imposed: Free boundary conditions are imposed at the end of the pipe at $z=L$ as usual by copying the values of the neighbouring cell normal to the surface. In addition, because the axis $r=0$ is now part of the simulation space (in contrast to the previous case, where the axis was a source boundary condition), we need to take care about the coordinates' singularity at $r=0$ by also implementing a free boundary condition at $r=r_{\rm min}=1$ (Fig.\ref{CCylinder}). Bounce back boundary conditions are considered in the pipe walls at $r=r_{\rm max}$ and in the closed end at $z=0$ by interchanging the distribution functions travelling on opposite velocity vectors. Finally, some dissipation must be added to reach steady-stable oscillations after a transient. This is done just by adding a damping factor for the distribution functions at the bounce-back boundary condition after the interaction with the walls,
\begin{equation}
f_{i_{\rm out}}\left(\vec{x}_{\rm wall},\vec{\xi}_{\rm out},t\right)=D\cdot f_{i_{\rm in}}\left(\vec{x}_{\rm wall},\vec{\xi}_{\rm in},t\right)\quad,
\end{equation}
where the indexes $in$ and $out$ means the incident and reflected directions, respectively. For all simulations we choose $D=0.65$, which was enough to reach stable normal oscillation modes after 8400 timesteps.

\begin{figure}
\centering
\includegraphics[width=.45\textwidth]{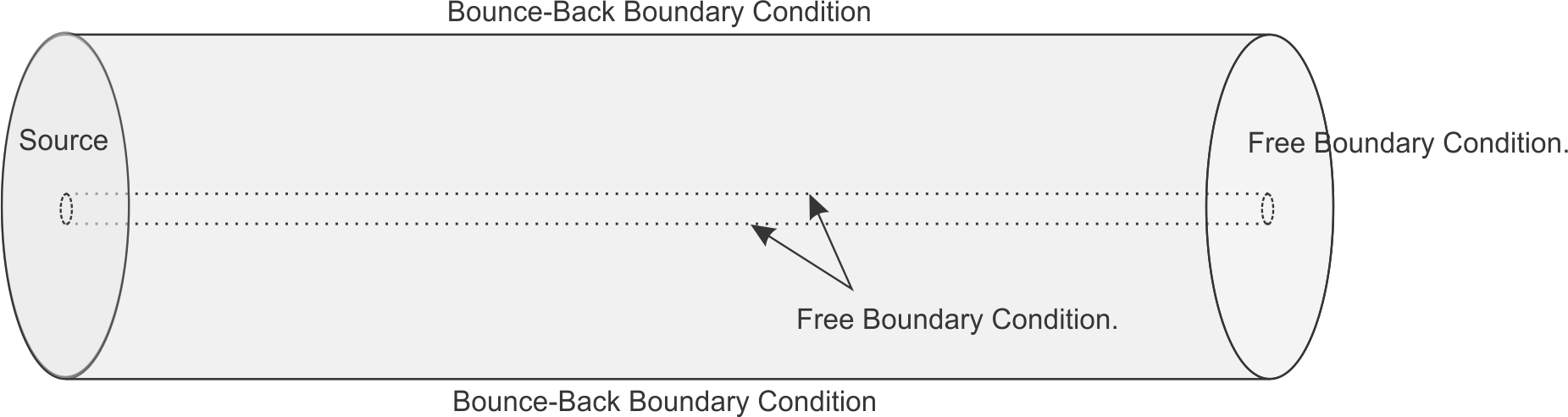}
\caption{Boundary conditions for the simulation of a wave inside a pipe}
\label{Fig.BoundariesPipe}
\end{figure}

As a first case, let us consider that the left end oscillates as a rigid body like a moving uniform plate; therefore, the functional form of the source is $
J'_z(r,\theta,0,t)=C\sin\left(\omega t\right)$. This wave propagates along the $z$ direction only (as in a one-dimensional cord) and, we could expect that the resonant frequencies correspond to the open-end pipe in one dimension given by $\omega_n=\frac{2\pi c(2n+1)}{4L_z}$. However, these frequencies are shifted and some peaks are suppressed (see Fig.\ref{Fig.Frequencies}) due to the appearance of radial modes that cannot be suppressed in this geometry, as will be shown below.
Next, we considered that the left end is a fixed wall plus an oscillating ring in front of it. The ring's radius is chosen to be of two cells, such that the ring fits just around the cylinder at $r=0$ that has been excluded from the simulation space. The functional form of this source is $P(r_{min},\theta,0,t)=C\sin(\omega t)$. The expected normal modes are now three-dimensional with axial symmetry. The pipe imposes a von Newmann boundary condition at the cylindrical wall,
\begin{equation}
\left(\frac{\partial}{\partial r} +\frac{1}{r} \right )P(r_{max},\theta,z)=0\quad .
\label{Boundary}
\end{equation}
Under such conditions the characteristic frequencies are (see Appendix \ref{TheoreticalPipe}):
\begin{equation}
\omega_{j,0,n}=c\left[\zeta_{l}^2+\left(\frac{(2n+1)\pi}{4z_{\rm max}} \right )^2 \right ]^{1/2}\quad .
\label{TheoreticalOmegaB}
\end{equation}

In order to identify the resonant frequencies for the two different perturbations, we vary the source frequency from $0$ to $0.03$ oscillations per timestep measuring the maximum pressure at a given point inside the pipe ($r=3,\, z=z_{\rm max}/10$). For these simulations, we set $c=0.1$. 
 Fig. \ref{Fig.Frequencies} shows the simulations results. The solid vertical lines are the expected values for the resonant frequencies obtained from Eq.(\ref{TheoreticalOmegaB}), the dashed line corresponds to the rigid uniform plate and the solid line, to the oscillating ring. Note that in the case of the uniform ring, longitudinal modes are more defined but shifted due to the appearance of the first radial mode at $\omega=0.164$, in addition, some resonant peaks that are present for the case of the ring were suppressed in the spectrum of the plate. The resonant frequencies simulated with the oscillating ring as source are in great agreement with the expected values, and the RMS deviation between the maxima and the black vertical lines is less than 1$\%$.  
\begin{figure}
\centering
\includegraphics[width=0.5\textwidth]{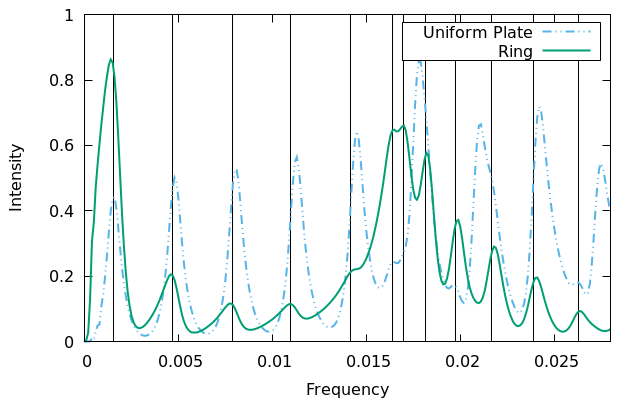} 
\caption{Pressure intensity at a fixed point inside a cylindrical pipe for waves generated by either an uniform plate or an oscillating ring. The solid vertical lines identify the theoretical frequencies for the normal modes.}
\label{Fig.Frequencies}
\end{figure}
Fig.\ref{PressureCylinder} shows a snapshot of the pressure waves inside the cylinder for the ring perturbation. 
\begin{figure}
\centering
\includegraphics[width=0.5\textwidth]{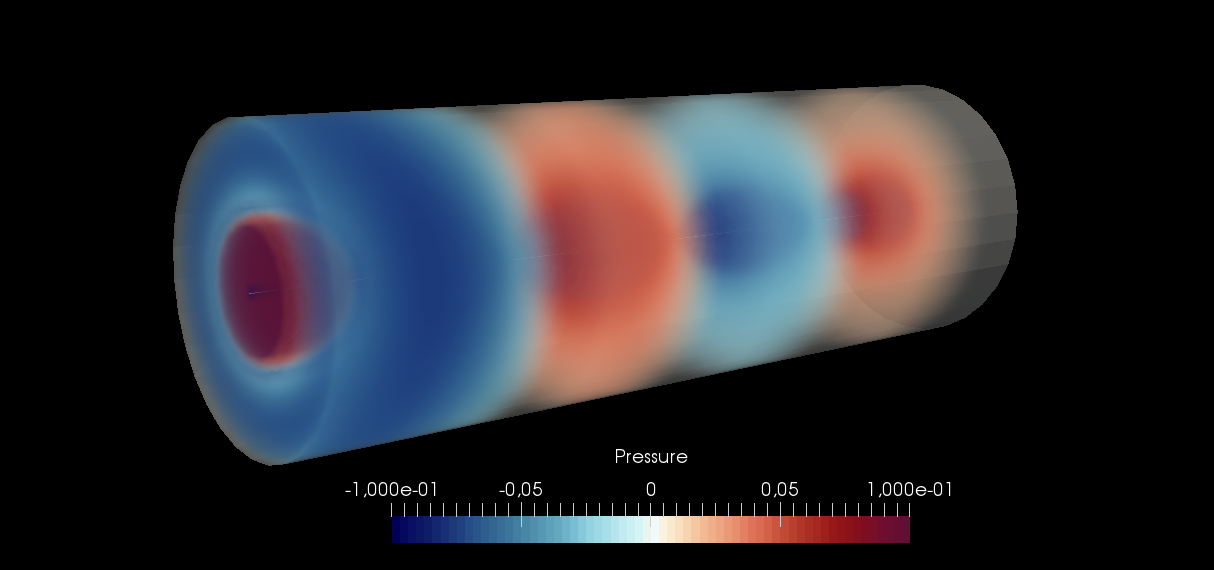} 
\caption{Pressure waves inside an open-closed end Pipe.}
\label{PressureCylinder}
\end{figure}

Next, let us measure the model convergence. To this aim, we choose a theoretically known quantity, in this case the third characteristic frequency of the open-closed end pipe, to be compared with its simulated value. As the resolution increases, we expect that the smaller the cell, the smaller the difference between the theoretical value and the simulated one. Such behaviour is shown in Fig. \ref{Fig.Convergence}. The error grows with cell size as a power law with exponent 2, that is the model convergence is second order. 

\begin{figure}
\centering
\includegraphics[width=0.5\textwidth]{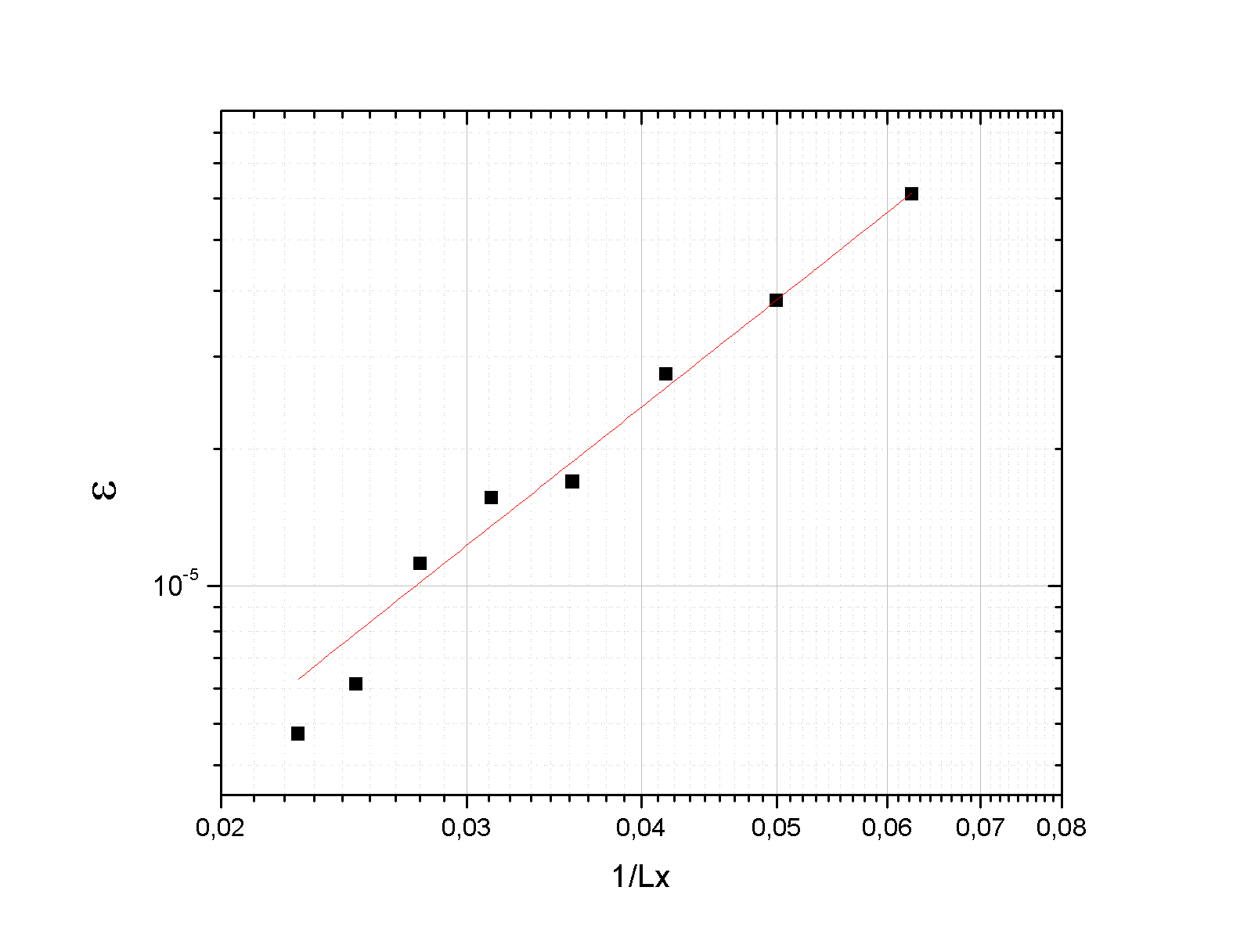} 
\caption{Simulation error for the third characteristic frequency of an closed-open end pipe ($\omega_{\rm theo}=0.00608$, for $n=3$, $z_{\rm max}=345.6$ and $c=0.5$ , see Eq.\ref{TheoreticalOmegaB}) as a function of the cell size. The red line is a power fit $\epsilon=A\left(1/L_x\right)^{B}$ with $A=0.012\pm 0.005$ and $B=2.105\pm 0.080$.}
\label{Fig.Convergence}
\end{figure}

\subsection{Trumpet}
The coordinate system Eq. \ref{cylindricalsystem} can be modified to include a flared end resembling a trumpet's bell. In this work, we consider a particular shape named \textit{Bessel Horn}, given by the following coordinate system
\begin{equation}
\begin{split}
x&=r\cos(\theta)\mathcal{Z}^{-\lambda}\\
y&=r\sin(\theta)\mathcal{Z}^{-\lambda}\\
z&=z\\
\end{split} \quad,
\label{coordinatesTrumpet}
\end{equation}
where $\lambda$ is a positive real number and $\mathcal{Z}=Z_\text{max}-z$ with $Z_\text{max}$ the trumpet's length (Fig.\ref{CTrumpet}).
\begin{figure}
\centering
\includegraphics[width=0.5\textwidth]{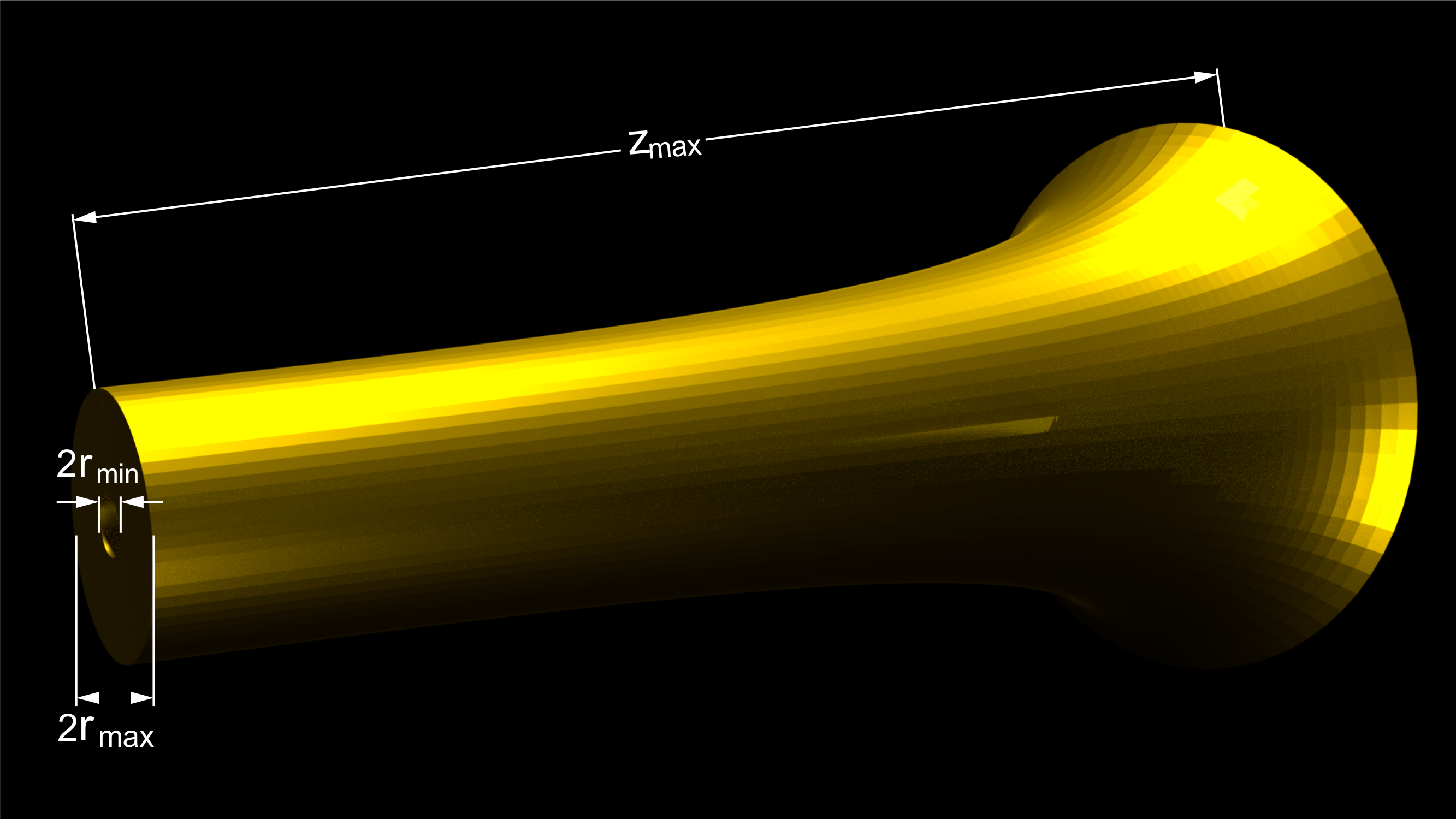} 
\caption{Coordinate transformation to simulate the trumpet.}
\label{CTrumpet}
\end{figure}
The surfaces for $r=\text{const}$ resembles the shape of a different horn for each value of $\lambda$ (Fig. \ref{Fig.ValuesofLambda}).\\
\begin{figure}
\centering
\includegraphics[width=0.5\textwidth]{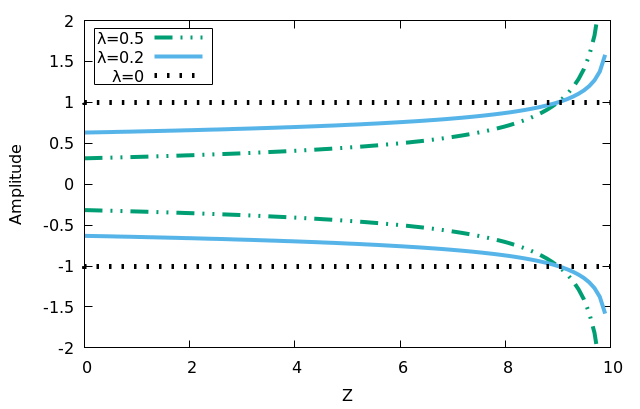} 
\caption{Shape of the trumpet for different values of $\lambda$. }
\label{Fig.ValuesofLambda}
\end{figure}

The Non-zero Christoffel symbols calculated for these coordinates in terms of $\lambda$ are given by
\begin{equation}
\begin{split}
\Gamma^r_{rz}=\frac{\lambda}{\mathcal{Z}}=\Gamma^r_{zr},\quad \Gamma^r_{zz}=\frac{r\lambda(\lambda+1)}{\mathcal{Z}^2}\\
\Gamma^\theta_{z\theta}=\frac{\lambda}{\mathcal{Z}^2}=\Gamma^\theta_{\theta z},\quad \Gamma^r_{\theta\theta}=-r,\quad  \Gamma^\theta_{r \theta}=\frac{1}{r}= \Gamma^\theta_{\theta r}\, ,
\end{split}
\label{ChristoffelTrumpet}
\end{equation}
and the metric tensor, by
\begin{equation}
g=\left( \begin {matrix}\mathcal{Z}^{-2\,\lambda}&0& \mathcal{Z}^{-2\,\lambda-1}\lambda\,r\\ 0& \mathcal{Z}^{-2\,\lambda}{r}^{2}&0\\ \mathcal{Z}^{-2\,\lambda-1}\lambda\,r&0&{\frac { \mathcal{Z}^{-2\,\lambda}{\lambda}^{2}{r}^{2}+ \mathcal{Z}^{2}}{ \mathcal{Z}^{2}}}\end {matrix}\right)\quad.
\label{MetricTrumpet}
\end{equation}

By using Eqs. \ref{coordinatesTrumpet}, \ref{ChristoffelTrumpet} and \ref{MetricTrumpet} it is possible to simulate the acoustical waves inside a trumpet. The simulation space and boundary conditions are the same to the ones for the pipe with an oscillating ring (Fig.~\ref{Fig.BoundariesPipe}). Fig.~\ref{Fig.Trumpet}  shows a snapshot after 8400 timesteps of the pressure profile for an axisymmetric Bessel horn with $\lambda=0.2$, $r_{min}=2$, $r_{max}=150$ and $Z_\text{max}=750$, at a frequency $\omega=18\pi c/4z_{\rm max}$. As expected, the intensity of the pressure wave is reduced at the end of the trumpet due to the bell.
Fig.~\ref{FigFrequenciesBessel} compares the oscillation envelope as a function of $z$ at $r=3$ cells when the ring oscillates at fixed frequency  $\omega=0.005$ for two shapes: a horn and a cylindrical pipe. Note that the amplitude of the oscillations decrease near the bell when it is present (at highest values of $z$ when $\lambda\neq 0$)
\begin{figure}
\centering
\includegraphics[width=0.5\textwidth]{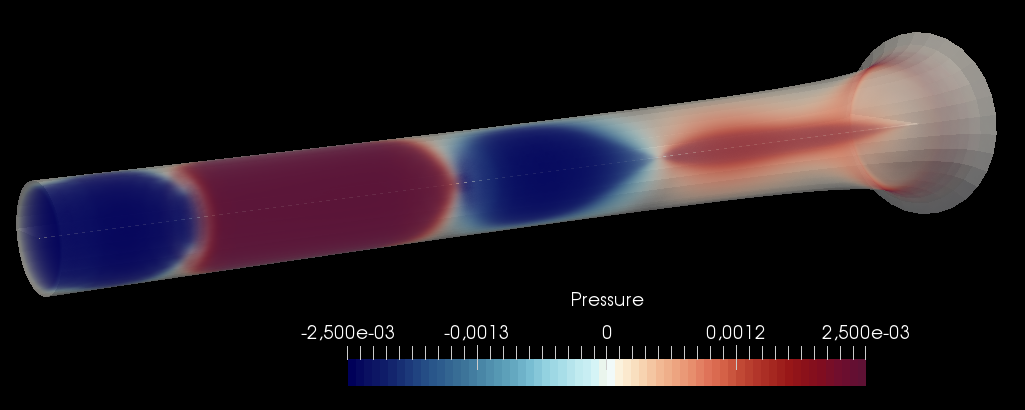} 
\caption{Steady-state pressure profile inside a trumpet at a frequency $\omega=18\pi c/4z_{\rm max}$.}
\label{Fig.Trumpet}
\end{figure}

\begin{figure}
\centering
\includegraphics[width=0.5\textwidth]{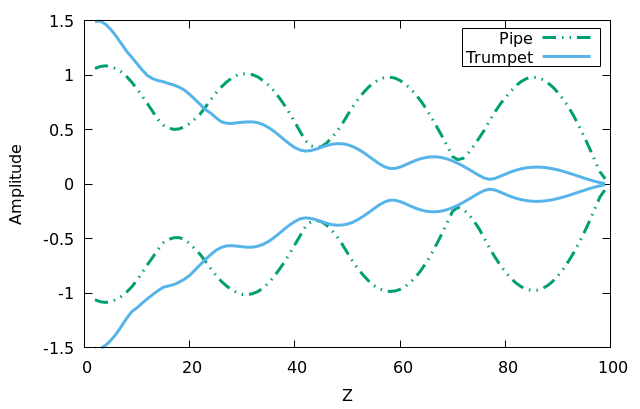} 
\caption{Envelope of oscillation for the pipe ($\lambda=0$) and the trumpet ($\lambda=0.1$)}
\label{FigFrequenciesBessel}
\end{figure}

Let us now study how the natural frequencies change by the inclusion of the bell. Figure \ref{Fig.TrumpetPipe} shows the resonant peaks for a trumpet with $\lambda=0.1$, compared with the previously obtained results for $\lambda=0$.
As expected, the longitudinal normal modes for the trumpet are less defined, and the second radial modes are strongly suppressed by the bell, similarly to what is observed in real pipes \cite{Rossing2002}.

\begin{figure}
\centering
\includegraphics[width=0.5\textwidth]{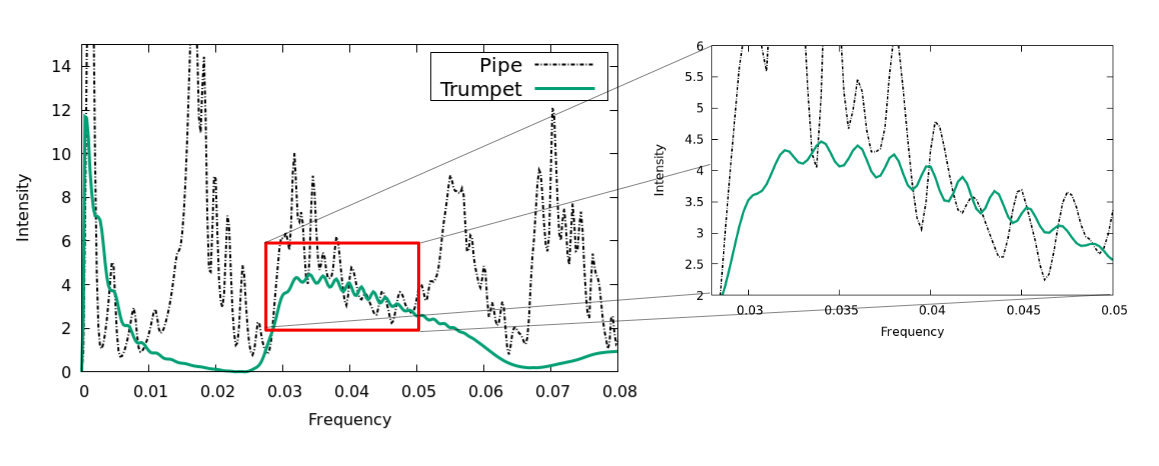} 
\caption{Comparison between the vibrational modes of a pipe and a trumpet with $\lambda=0.1$ and zoom over the red square.}
\label{Fig.TrumpetPipe}
\end{figure}

\subsection{Torus}
As a third benchmark, we study the pressure waves inside a torus. A torus can be seen as a circle of radius $r_{\rm max}$ whose center travels along a larger circle of radius $R$ (See Fig.  \ref{CTorus}). It can be described by the coordinate transformation
\begin{equation}
\begin{split}
x&=\left(R+r\cos(\phi)\right)\cos(\theta)\\
y&=\left(R+r\cos(\phi)\right)\sin(\theta)\\
z&=r\sin(\phi)\\
\end{split} \quad.
\label{CoordinatesTorus}
\end{equation}
The metric tensor for this coordinates is, therefore,
\begin{equation}
g=\left( \begin {matrix}1&0&0\\ 0& \left(R+r\cos(\phi)\right)^2&0\\ 0&0&r^2\end{matrix}\right)\quad,
\label{AMetricTorus}
\end{equation}
and the non-zero Christoffel symbols are
\begin{equation}
\begin{split}
\Gamma^r_{\theta\theta}&=\left(R+r\cos(\phi)\right)\cos(\phi),\quad \Gamma^r_{\phi\phi}=\left(-r\right),\,\\
\Gamma^\theta_{r\theta}&=\frac{\cos(\phi)}{\left(R+r\cos(\phi)\right)}=\Gamma^\theta_{\theta r},\quad \Gamma^\theta_{\theta\phi}=\frac{-\sin(\phi)r}{\left(R+r\cos(\phi)\right)}=\Gamma^\theta_{\phi\theta},\quad\\
\Gamma^\phi_{r \phi}&=\frac{1}{r}= \Gamma^\phi_{\phi r},\, \Gamma_\phi^{\theta\theta}=\frac{\left(R+r\cos(\phi)\right)\sin(\phi)}{r}\, .
\end{split}
\label{AChristoffelTorus}
\end{equation}
\begin{figure}
\centering
\includegraphics[width=0.40\textwidth]{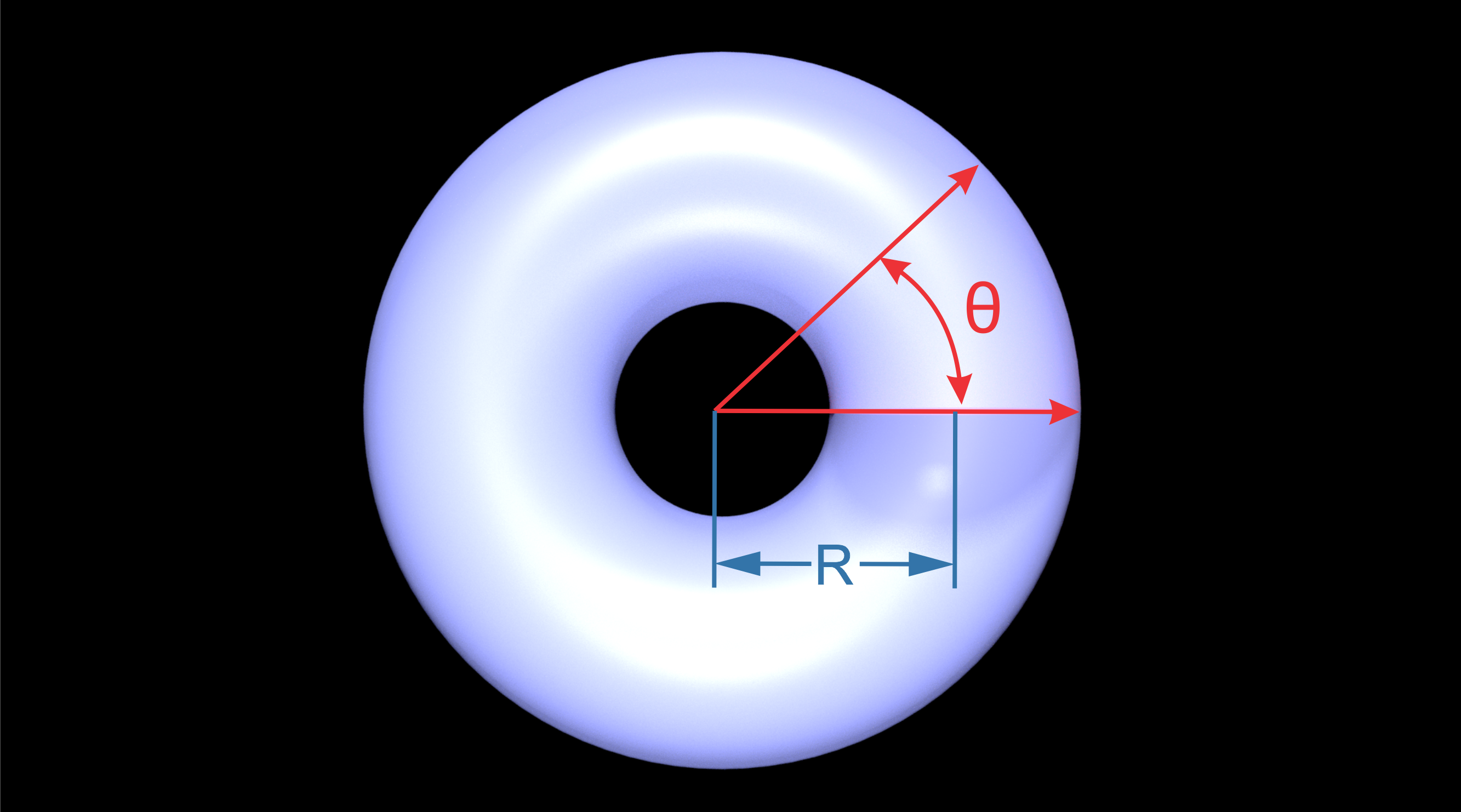} 
\includegraphics[width=0.40\textwidth]{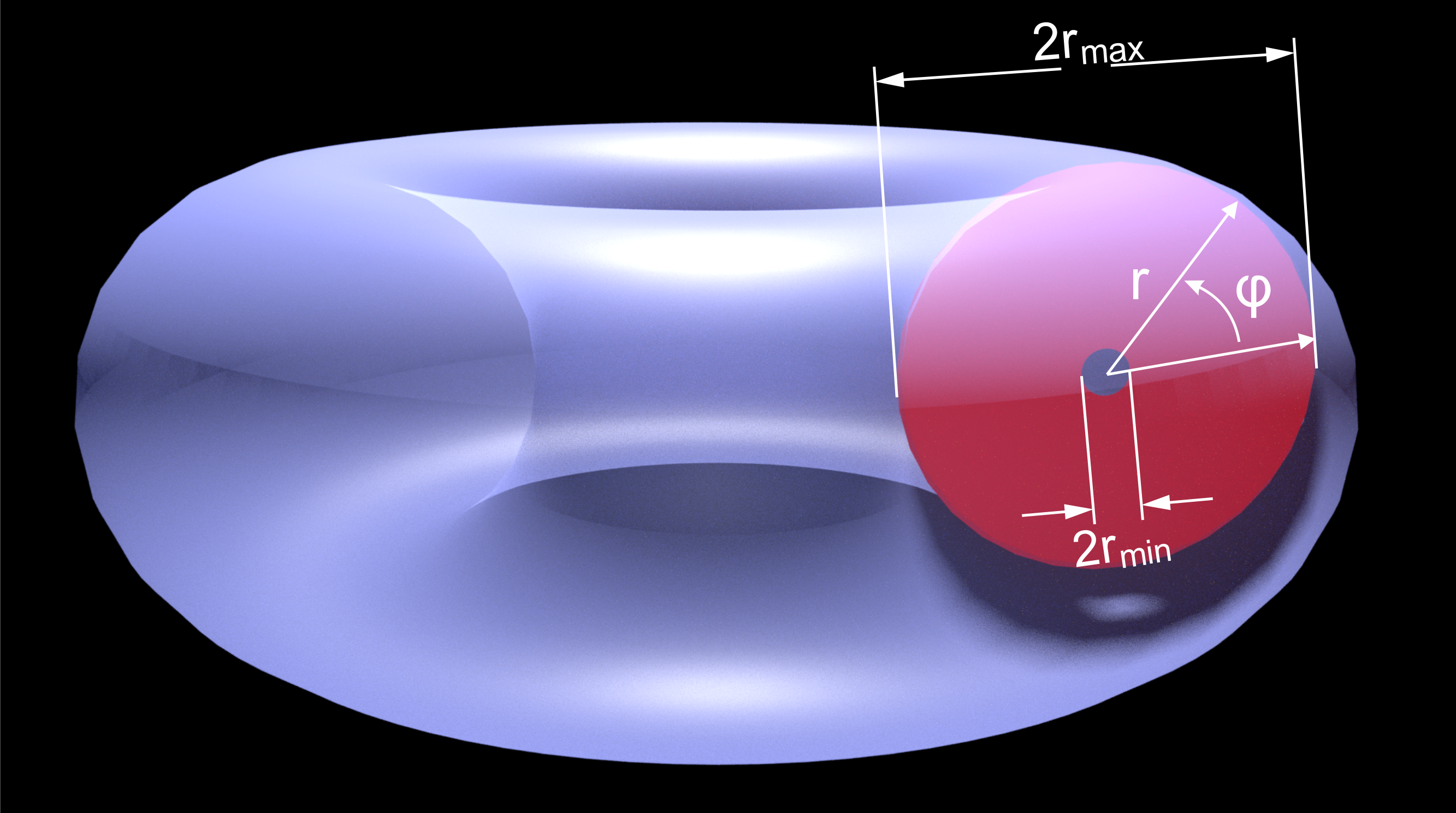} 
\caption{Coordinate transformation and definition for the Torus geometry. }
\label{CTorus}
\end{figure}
The chosen torus has $R=640$, $5.12<r<409.6$, $0<\theta<2\pi$ and $0<\phi<2\pi$. For the simulations we used 20 cells in $r$, 200 cells in $\theta$ and 30 cells in $\phi$. The boundary conditions imposed are bounce-back with damping factor $D=0.65$ in $r_{\rm max}$, free in $r_{\rm min}$ and periodic both in $\theta$ and $\phi$. 

To test the LBM we run three different simulations, each one with a different source. The first one is a ring along the major circle of radius $R$, and its functional form is $P_{source}(r,\theta,\phi,t)=A\sin(\omega t)\delta(r-r_{min})$. Fig. \ref{ToroRadial1} shows the evolution of the pressure wave, showing expected radial symmetry.
\begin{figure}
\centering
\includegraphics[width=0.40\textwidth]{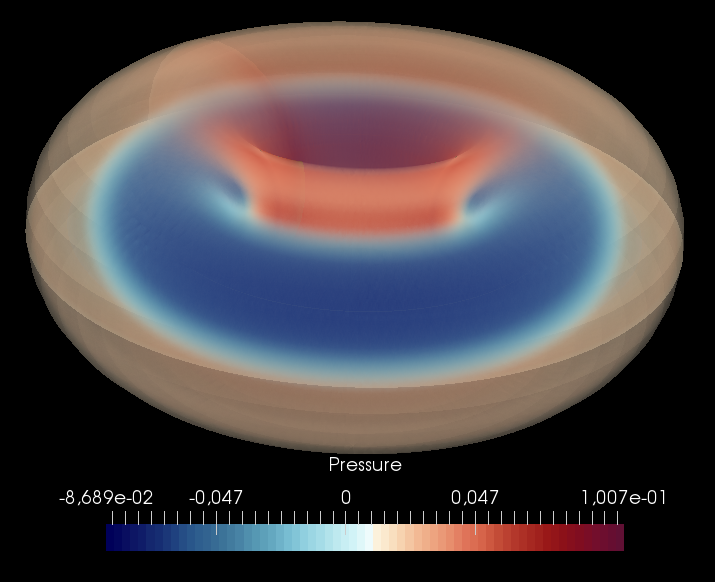} 
\caption{Pressure waves inside a torus. The waves were generated from the inner torus with $r=r_{\rm min}$ and for all the values of $\theta$ and $\phi$.}
\label{ToroRadial1}
\end{figure}
 The second source is similar to the small ring employed in pipes and trumpets, with functional form $P_{source}(r,\theta,\phi,t)=A\sin(\omega t)\delta(r-r_{\rm min})\delta(\theta)$. This source allows the excitation of normal modes in both $r$ and $\theta$ directions. The results for different frequencies are shown in Fig. \ref{ToroAngular}.
\begin{figure}
\centering
\includegraphics[width=0.2\textwidth]{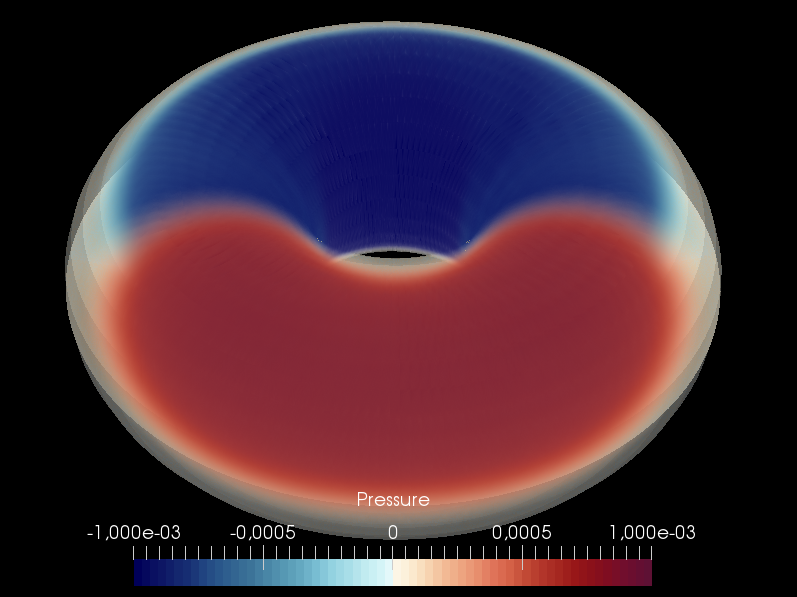} 
\includegraphics[width=0.2\textwidth]{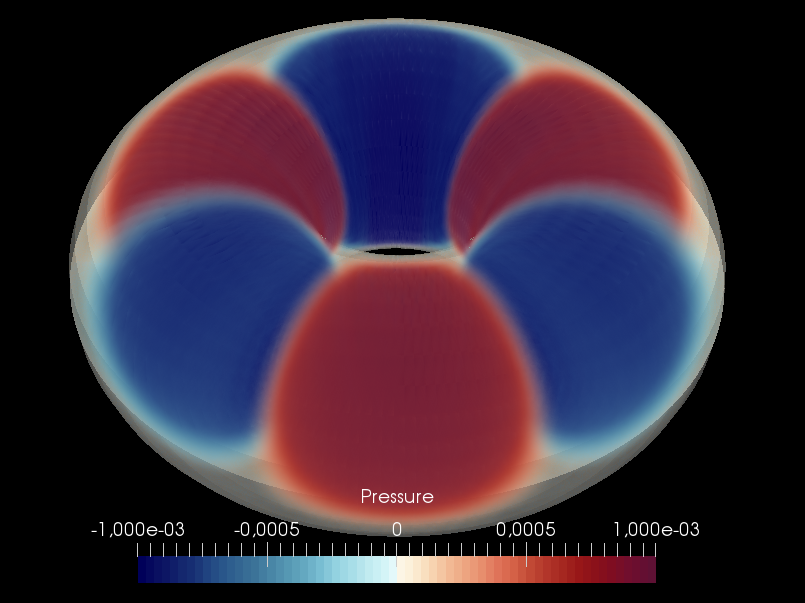} 
\includegraphics[width=0.2\textwidth]{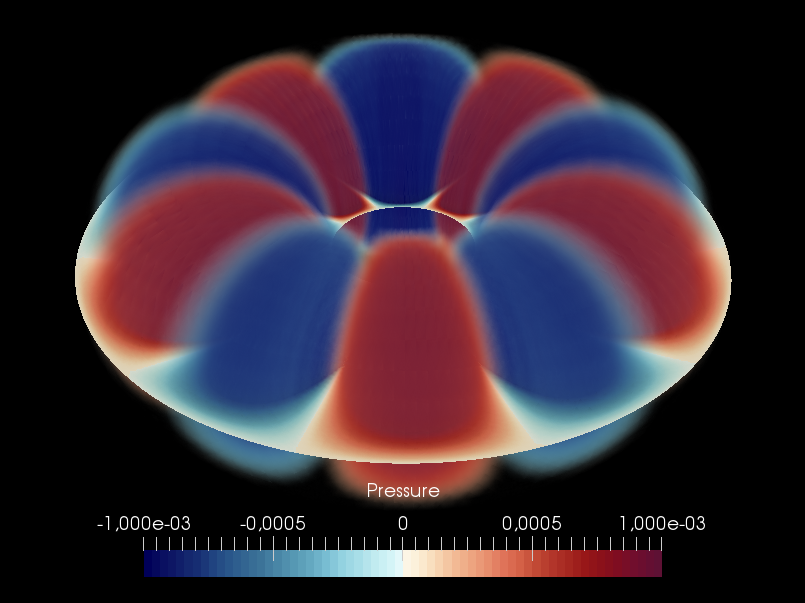} 
\includegraphics[width=0.2\textwidth]{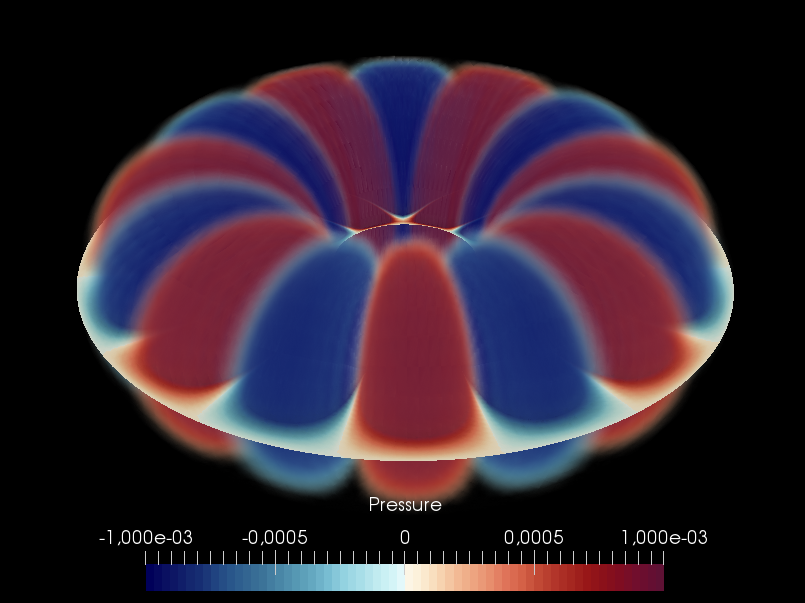} 
\caption{Pressure waves inside a torus. The waves were generated from a ring with $r=r_{min}$ and $\theta=0$ for all the values of $\phi$, the wavelengths $\lambda$ are $2\pi R/n$ where $n=2,\, 4,\, 9,\, 19$ (left to right and top to bottom). The frequency $\omega=2\pi c/\lambda$}
\label{ToroAngular}
\end{figure}
We also studied the resonant frequencies for this geometry. To this aim we located the small oscillating ring at $\theta=0$ and measured the pressure intensity at the opposite point on the major circle (i. e. at $\theta=\pi$) after 8000 time steps, when the wave has reached a steady-stable condition. Fig. \ref{FreqTorus} shows the obtained intensity for frequencies between 0 and 0.5 radians per click varying the frequency in steps of $\Delta \omega = 0.001$. 
\begin{figure}
\centering
\includegraphics[width=0.5\textwidth]{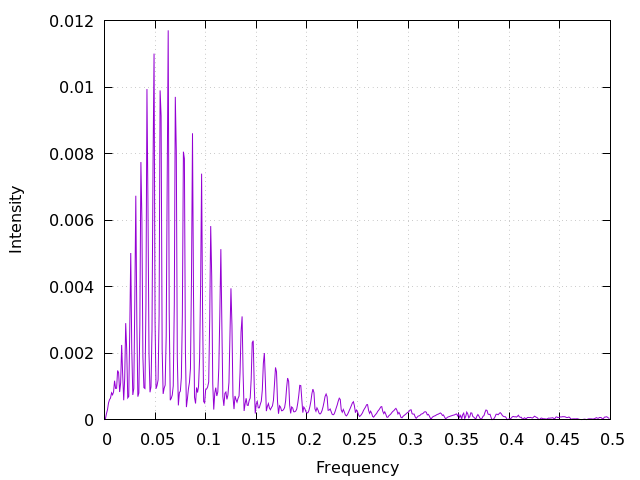} 
\caption{Wave intensity at $(r_{max},\pi,0)$ after $8000$ timesteps for $0<\omega<0.5$ with $\Delta \omega = 0.001$ }
\label{FreqTorus}
\end{figure}
Note that the larger resonant peaks are placed whitin 0.05 and 0.1 radians per click, with a cutoff frequency near to 0.25 radians per click. Although the analytical solution for the normal modes of the torus was not found in the literature, it is interesting to point out that the pattern is very similar to the one for a trumpet with an open bell \cite{Rossing2002}.

To simulate more general perturbations, the third source we used is an oscillating point outside any symmetry axis, i.e. at $r=\frac{r_{max}-r_{min})}{2}$, $\theta=0$, and $\phi=\pi/2$. The functional form of the source is, therefore,  $P_{source}(r,\theta,\phi)=A\sin(\omega t)\delta\left(r-\frac{r_{max}-r_{min})}{2}\right)\delta(\theta)\delta(\phi-\pi/2)$. The resulting pattern is shown in Fig. \ref{ToroAsimetrico}. We can observe that the pattern is no longer symmetric in $\phi$ and shows variations along $r$. 
\begin{figure}
\centering
\includegraphics[width=0.4\textwidth]{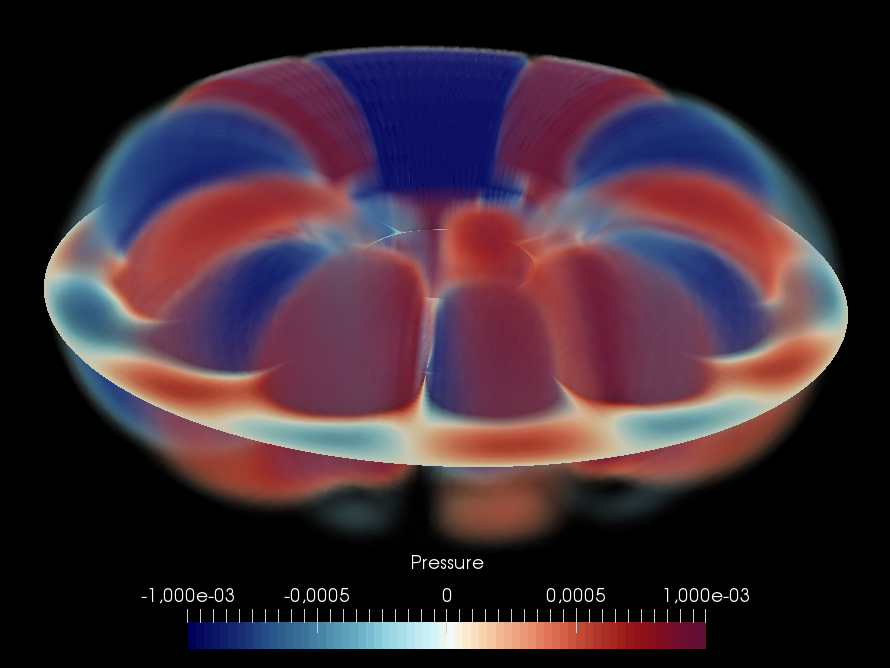} 
\caption{General profile of the pressure waves inside the torus. }
\label{ToroAsimetrico}
\end{figure}
For this last case, the evolution inside the torus is general, since it is not restricted to any type of symmetry, showing that our LBM can be applied to non-axisymmetric systems.

\section{Conclusions}
\label{Conclusions}
A lattice-Boltzmann model (LBM) for the simulation of acoustic waves in general curvilinear coordinates was developed and implemented. The model considers the additional terms related to the wave equation in curvilinear coordinates as forcing terms of the conservation laws and uses the approach of Guo \textit{et. al.} to construct the proper forced lattice-Boltzmann model.
The model was tested by reproducing the propagation of a cylindrical wave, by finding the normal modes inside a cylindrical pipe and a Bessel horn trumpet and by computing steady-state pressure profiles inside a Torus. Free boundary conditions were set at the open ends, and bounce-back boundary conditions at the walls, with a damping factor in the lateral ones. The simulation results show  very good agreement with the theoretical expectations: both the RMS errors for the cylindrical wave and the resonant frequencies for the pipe are less than 1$\%$. In addition, the results for the trumpet show a strong suppression of the second radial modes, similar to the one observed in real trumpets. For the case of the Torus, the resonant spectrum shows an interesting similarity with the one for a trumpet and, although we did not find theoretical expressions for the normal modes, our results are in qualitative agreement with the ones expected for that geometry.

Unlike previous models, our proposed LBM is completely general and does not ask for additional interpolation steps. The geometric information of the coordinate system is completely included in the equilibrium distribution function and in the macroscopic quantities, but the system stored in the computer is still a cubic lattice. For this reason, it is not necessary a new discretization scheme, i.e. the discrete velocity set is defined as usual and the velocity vectors are the same for all cells. The model also allows to take advantage of the symmetry of the system to reduce the number of cells to reach certain precision, reducing computational requirements. For instance, our simulations of pipes and horns are actually two-dimensional in cylindrical coordinates, taking advantage of the azimutal symmetry, and therefore, they consume fewer computational resources. Those versatilities widely opens the application range of lattice-Boltzmann models to acoustic systems with very complicated geometries, like more complex musical instruments, auditoriums and concert halls or even detection organs as the Cochlea. The convergence is second order, which keeps the accuracy of the standard lattice-Boltzmann models.

Summarizing, this work introduces a lattice-Boltzmann model for the simulation of acoustic waves in general curvilinear coordinates. The manuscript has written in such a way that the interested reader can implement any desired geometry without effort, just by computing the metric tensor and Christoffel symbols and inserting them into the model. So, hopefully, it will be of great utility in future research.\\

\bibliography{BibliMSc}

\begin{thebibliography}{28}%
\makeatletter
\providecommand \@ifxundefined [1]{%
 \@ifx{#1\undefined}
}%
\providecommand \@ifnum [1]{%
 \ifnum #1\expandafter \@firstoftwo
 \else \expandafter \@secondoftwo
 \fi
}%
\providecommand \@ifx [1]{%
 \ifx #1\expandafter \@firstoftwo
 \else \expandafter \@secondoftwo
 \fi
}%
\providecommand \natexlab [1]{#1}%
\providecommand \enquote  [1]{``#1''}%
\providecommand \bibnamefont  [1]{#1}%
\providecommand \bibfnamefont [1]{#1}%
\providecommand \citenamefont [1]{#1}%
\providecommand \href@noop [0]{\@secondoftwo}%
\providecommand \href [0]{\begingroup \@sanitize@url \@href}%
\providecommand \@href[1]{\@@startlink{#1}\@@href}%
\providecommand \@@href[1]{\endgroup#1\@@endlink}%
\providecommand \@sanitize@url [0]{\catcode `\\12\catcode `\$12\catcode
  `\&12\catcode `\#12\catcode `\^12\catcode `\_12\catcode `\%12\relax}%
\providecommand \@@startlink[1]{}%
\providecommand \@@endlink[0]{}%
\providecommand \url  [0]{\begingroup\@sanitize@url \@url }%
\providecommand \@url [1]{\endgroup\@href {#1}{\urlprefix }}%
\providecommand \urlprefix  [0]{URL }%
\providecommand \Eprint [0]{\href }%
\providecommand \doibase [0]{http://dx.doi.org/}%
\providecommand \selectlanguage [0]{\@gobble}%
\providecommand \bibinfo  [0]{\@secondoftwo}%
\providecommand \bibfield  [0]{\@secondoftwo}%
\providecommand \translation [1]{[#1]}%
\providecommand \BibitemOpen [0]{}%
\providecommand \bibitemStop [0]{}%
\providecommand \bibitemNoStop [0]{.\EOS\space}%
\providecommand \EOS [0]{\spacefactor3000\relax}%
\providecommand \BibitemShut  [1]{\csname bibitem#1\endcsname}%
\let\auto@bib@innerbib\@empty
\bibitem [{\citenamefont {Mendoza}(2012)}]{MENDOZA}%
  \BibitemOpen
  \bibfield  {author} {\bibinfo {author} {\bibfnamefont {M.}~\bibnamefont
  {Mendoza}},\ }\href@noop {} {\enquote {\bibinfo {title} {{Relativistic Fluid
  Dynamics in Complex Systems}},}\ } (\bibinfo {year} {2012})\BibitemShut
  {NoStop}%
\bibitem [{\citenamefont {Succi}(2001)}]{Succi2001}%
  \BibitemOpen
  \bibfield  {author} {\bibinfo {author} {\bibfnamefont {S.}~\bibnamefont
  {Succi}},\ }\href@noop {} {\emph {\bibinfo {title} {{The lattice Boltzmann
  equation for fluid dynamics and beyond}}}}\ (\bibinfo  {publisher} {Clarendon
  Press},\ \bibinfo {year} {2001})\ p.\ \bibinfo {pages} {288}\BibitemShut
  {NoStop}%
\bibitem [{\citenamefont {Chopard}\ \emph {et~al.}(1997)\citenamefont
  {Chopard}, \citenamefont {Luthi},\ and\ \citenamefont {Wagen}}]{Chopard1997}%
  \BibitemOpen
  \bibfield  {author} {\bibinfo {author} {\bibfnamefont {B.}~\bibnamefont
  {Chopard}}, \bibinfo {author} {\bibfnamefont {P.}~\bibnamefont {Luthi}}, \
  and\ \bibinfo {author} {\bibfnamefont {J.-F.}\ \bibnamefont {Wagen}},\ }\href
  {\doibase 10.1049/ip-map:19971197} {\bibfield  {journal} {\bibinfo  {journal}
  {IEE Proceedings - Microwaves, Antennas and Propagation}\ }\textbf {\bibinfo
  {volume} {144}},\ \bibinfo {pages} {251} (\bibinfo {year}
  {1997})}\BibitemShut {NoStop}%
\bibitem [{\citenamefont {Chopard}(2009)}]{Chopard2009}%
  \BibitemOpen
  \bibfield  {author} {\bibinfo {author} {\bibfnamefont {B.}~\bibnamefont
  {Chopard}},\ }in\ \href {\doibase 10.1007/978-0-387-30440-3_57} {\emph
  {\bibinfo {booktitle} {Encyclopedia of Complexity and Systems Science}}}\
  (\bibinfo  {publisher} {Springer New York},\ \bibinfo {address} {New York,
  NY},\ \bibinfo {year} {2009})\ pp.\ \bibinfo {pages} {865--892}\BibitemShut
  {NoStop}%
\bibitem [{\citenamefont {Guangwu}(2000)}]{Guangwu2000}%
  \BibitemOpen
  \bibfield  {author} {\bibinfo {author} {\bibfnamefont {Y.}~\bibnamefont
  {Guangwu}},\ }\href {\doibase 10.1006/jcph.2000.6486} {\bibfield  {journal}
  {\bibinfo  {journal} {Journal of Computational Physics}\ }\textbf {\bibinfo
  {volume} {161}},\ \bibinfo {pages} {61} (\bibinfo {year} {2000})}\BibitemShut
  {NoStop}%
\bibitem [{\citenamefont {Mendoza}\ and\ \citenamefont
  {Mu{\~{n}}oz}(2010)}]{Mendoza2010}%
  \BibitemOpen
  \bibfield  {author} {\bibinfo {author} {\bibfnamefont {M.}~\bibnamefont
  {Mendoza}}\ and\ \bibinfo {author} {\bibfnamefont {J.~D.}\ \bibnamefont
  {Mu{\~{n}}oz}},\ }\href {\doibase 10.1103/PhysRevE.82.056708} {\bibfield
  {journal} {\bibinfo  {journal} {Physical Review E}\ }\textbf {\bibinfo
  {volume} {82}},\ \bibinfo {pages} {056708} (\bibinfo {year}
  {2010})}\BibitemShut {NoStop}%
\bibitem [{\citenamefont {Palpacelli}\ and\ \citenamefont
  {Succi}(2008)}]{Palpacelli2008}%
  \BibitemOpen
  \bibfield  {author} {\bibinfo {author} {\bibfnamefont {S.}~\bibnamefont
  {Palpacelli}}\ and\ \bibinfo {author} {\bibfnamefont {S.}~\bibnamefont
  {Succi}},\ }\href {\doibase 10.1103/PhysRevE.77.066708} {\bibfield  {journal}
  {\bibinfo  {journal} {Physical Review E}\ }\textbf {\bibinfo {volume} {77}},\
  \bibinfo {pages} {066708} (\bibinfo {year} {2008})}\BibitemShut {NoStop}%
\bibitem [{\citenamefont {Li}\ and\ \citenamefont {Shan}(2011)}]{Li2011}%
  \BibitemOpen
  \bibfield  {author} {\bibinfo {author} {\bibfnamefont {Y.}~\bibnamefont
  {Li}}\ and\ \bibinfo {author} {\bibfnamefont {X.}~\bibnamefont {Shan}},\
  }\href {\doibase 10.1098/rsta.2011.0109} {\bibfield  {journal} {\bibinfo
  {journal} {Philosophical transactions. Series A, Mathematical, physical, and
  engineering sciences}\ }\textbf {\bibinfo {volume} {369}},\ \bibinfo {pages}
  {2371} (\bibinfo {year} {2011})}\BibitemShut {NoStop}%
\bibitem [{\citenamefont {Sun}\ \emph {et~al.}(2015)\citenamefont {Sun},
  \citenamefont {P{\'{e}}rot}, \citenamefont {Zhang}, \citenamefont {Lew},
  \citenamefont {Mann}, \citenamefont {Gupta}, \citenamefont {Freed},
  \citenamefont {Staroselsky},\ and\ \citenamefont {Chen}}]{Sun2015}%
  \BibitemOpen
  \bibfield  {author} {\bibinfo {author} {\bibfnamefont {C.}~\bibnamefont
  {Sun}}, \bibinfo {author} {\bibfnamefont {F.}~\bibnamefont {P{\'{e}}rot}},
  \bibinfo {author} {\bibfnamefont {R.}~\bibnamefont {Zhang}}, \bibinfo
  {author} {\bibfnamefont {P.~T.}\ \bibnamefont {Lew}}, \bibinfo {author}
  {\bibfnamefont {A.}~\bibnamefont {Mann}}, \bibinfo {author} {\bibfnamefont
  {V.}~\bibnamefont {Gupta}}, \bibinfo {author} {\bibfnamefont {D.~M.}\
  \bibnamefont {Freed}}, \bibinfo {author} {\bibfnamefont {I.}~\bibnamefont
  {Staroselsky}}, \ and\ \bibinfo {author} {\bibfnamefont {H.}~\bibnamefont
  {Chen}},\ }\href {\doibase 10.1016/j.crme.2015.07.013} {\enquote {\bibinfo
  {title} {{Lattice Boltzmann formulation for flows with acoustic porous
  media}},}\ } (\bibinfo {year} {2015})\BibitemShut {NoStop}%
\bibitem [{\citenamefont {Salomons}\ \emph {et~al.}(2016)\citenamefont
  {Salomons}, \citenamefont {Lohman},\ and\ \citenamefont
  {Zhou}}]{Salomons2016}%
  \BibitemOpen
  \bibfield  {author} {\bibinfo {author} {\bibfnamefont {E.~M.}\ \bibnamefont
  {Salomons}}, \bibinfo {author} {\bibfnamefont {W.~J.~A.}\ \bibnamefont
  {Lohman}}, \ and\ \bibinfo {author} {\bibfnamefont {H.}~\bibnamefont
  {Zhou}},\ }\href {\doibase 10.1371/journal.pone.0147206} {\bibfield
  {journal} {\bibinfo  {journal} {PLOS ONE}\ }\textbf {\bibinfo {volume}
  {11}},\ \bibinfo {pages} {e0147206} (\bibinfo {year} {2016})}\BibitemShut
  {NoStop}%
\bibitem [{\citenamefont {Nannelli}\ and\ \citenamefont
  {Succi}(1992)}]{Nannelli1992}%
  \BibitemOpen
  \bibfield  {author} {\bibinfo {author} {\bibfnamefont {F.}~\bibnamefont
  {Nannelli}}\ and\ \bibinfo {author} {\bibfnamefont {S.}~\bibnamefont
  {Succi}},\ }\href {\doibase 10.1007/BF01341755} {\bibfield  {journal}
  {\bibinfo  {journal} {Journal of Statistical Physics}\ }\textbf {\bibinfo
  {volume} {68}},\ \bibinfo {pages} {401} (\bibinfo {year} {1992})}\BibitemShut
  {NoStop}%
\bibitem [{\citenamefont {He}\ and\ \citenamefont {Doolen}(1997)}]{He1997}%
  \BibitemOpen
  \bibfield  {author} {\bibinfo {author} {\bibfnamefont {X.}~\bibnamefont
  {He}}\ and\ \bibinfo {author} {\bibfnamefont {G.~D.}\ \bibnamefont
  {Doolen}},\ }\href {\doibase 10.1103/PhysRevE.56.434} {\bibfield  {journal}
  {\bibinfo  {journal} {Physical Review E}\ }\textbf {\bibinfo {volume} {56}},\
  \bibinfo {pages} {434} (\bibinfo {year} {1997})}\BibitemShut {NoStop}%
\bibitem [{\citenamefont {He}\ \emph {et~al.}(1996)\citenamefont {He},
  \citenamefont {Luo},\ and\ \citenamefont {Dembo}}]{He1996}%
  \BibitemOpen
  \bibfield  {author} {\bibinfo {author} {\bibfnamefont {X.}~\bibnamefont
  {He}}, \bibinfo {author} {\bibfnamefont {L.-S.}\ \bibnamefont {Luo}}, \ and\
  \bibinfo {author} {\bibfnamefont {M.}~\bibnamefont {Dembo}},\ }\href
  {\doibase 10.1006/JCPH.1996.0255} {\bibfield  {journal} {\bibinfo  {journal}
  {Journal of Computational Physics}\ }\textbf {\bibinfo {volume} {129}},\
  \bibinfo {pages} {357} (\bibinfo {year} {1996})}\BibitemShut {NoStop}%
\bibitem [{\citenamefont {Filippova}\ and\ \citenamefont
  {H{\"{a}}nel}(1998)}]{Filippova1998}%
  \BibitemOpen
  \bibfield  {author} {\bibinfo {author} {\bibfnamefont {O.}~\bibnamefont
  {Filippova}}\ and\ \bibinfo {author} {\bibfnamefont {D.}~\bibnamefont
  {H{\"{a}}nel}},\ }\href {\doibase 10.1006/jcph.1998.6089} {\bibfield
  {journal} {\bibinfo  {journal} {Journal of Computational Physics}\ }\textbf
  {\bibinfo {volume} {147}},\ \bibinfo {pages} {219} (\bibinfo {year}
  {1998})}\BibitemShut {NoStop}%
\bibitem [{\citenamefont {Li}\ \emph {et~al.}(2010)\citenamefont {Li},
  \citenamefont {He}, \citenamefont {Tang},\ and\ \citenamefont
  {Tao}}]{Li2010}%
  \BibitemOpen
  \bibfield  {author} {\bibinfo {author} {\bibfnamefont {Q.}~\bibnamefont
  {Li}}, \bibinfo {author} {\bibfnamefont {Y.~L.}\ \bibnamefont {He}}, \bibinfo
  {author} {\bibfnamefont {G.~H.}\ \bibnamefont {Tang}}, \ and\ \bibinfo
  {author} {\bibfnamefont {W.~Q.}\ \bibnamefont {Tao}},\ }\href {\doibase
  10.1103/PhysRevE.81.056707} {\bibfield  {journal} {\bibinfo  {journal}
  {Physical Review E - Statistical, Nonlinear, and Soft Matter Physics}\ }
  (\bibinfo {year} {2010}),\ 10.1103/PhysRevE.81.056707},\ \Eprint
  {http://arxiv.org/abs/1003.0282} {arXiv:1003.0282} \BibitemShut {NoStop}%
\bibitem [{\citenamefont {Reijers}\ \emph {et~al.}(2016)\citenamefont
  {Reijers}, \citenamefont {Gelderblom},\ and\ \citenamefont
  {Toschi}}]{Reijers2016}%
  \BibitemOpen
  \bibfield  {author} {\bibinfo {author} {\bibfnamefont {S.~A.}\ \bibnamefont
  {Reijers}}, \bibinfo {author} {\bibfnamefont {H.}~\bibnamefont {Gelderblom}},
  \ and\ \bibinfo {author} {\bibfnamefont {F.}~\bibnamefont {Toschi}},\ }\href
  {\doibase 10.1016/j.jocs.2016.05.012} {\bibfield  {journal} {\bibinfo
  {journal} {Journal of Computational Science}\ } (\bibinfo {year} {2016}),\
  10.1016/j.jocs.2016.05.012},\ \Eprint
  {http://arxiv.org/abs/arXiv:1512.01977v1} {arXiv:arXiv:1512.01977v1}
  \BibitemShut {NoStop}%
\bibitem [{\citenamefont {Budinski}(2012)}]{Budinski2012}%
  \BibitemOpen
  \bibfield  {author} {\bibinfo {author} {\bibfnamefont {L.}~\bibnamefont
  {Budinski}},\ }\href {\doibase 10.2166/hydro.2012.097} {\bibfield  {journal}
  {\bibinfo  {journal} {Journal of Hydroinformatics}\ } (\bibinfo {year}
  {2012}),\ 10.2166/hydro.2012.097}\BibitemShut {NoStop}%
\bibitem [{\citenamefont {Guo}\ \emph {et~al.}(2002)\citenamefont {Guo},
  \citenamefont {Zheng},\ and\ \citenamefont {Shi}}]{Guo2002}%
  \BibitemOpen
  \bibfield  {author} {\bibinfo {author} {\bibfnamefont {Z.}~\bibnamefont
  {Guo}}, \bibinfo {author} {\bibfnamefont {C.}~\bibnamefont {Zheng}}, \ and\
  \bibinfo {author} {\bibfnamefont {B.}~\bibnamefont {Shi}},\ }\href {\doibase
  10.1103/PhysRevE.65.046308} {\bibfield  {journal} {\bibinfo  {journal}
  {Physical Review E}\ }\textbf {\bibinfo {volume} {65}},\ \bibinfo {pages}
  {046308} (\bibinfo {year} {2002})}\BibitemShut {NoStop}%
\bibitem [{\citenamefont {Debus}\ \emph {et~al.}(2014)\citenamefont {Debus},
  \citenamefont {Mendoza},\ and\ \citenamefont {Herrmann}}]{Debus2014}%
  \BibitemOpen
  \bibfield  {author} {\bibinfo {author} {\bibfnamefont {J.-D.}\ \bibnamefont
  {Debus}}, \bibinfo {author} {\bibfnamefont {M.}~\bibnamefont {Mendoza}}, \
  and\ \bibinfo {author} {\bibfnamefont {H.~J.}\ \bibnamefont {Herrmann}},\
  }\href {\doibase 10.1103/PhysRevE.90.053308} {\bibfield  {journal} {\bibinfo
  {journal} {Physical Review E}\ }\textbf {\bibinfo {volume} {90}},\ \bibinfo
  {pages} {053308} (\bibinfo {year} {2014})}\BibitemShut {NoStop}%
\bibitem [{\citenamefont {Mendoza}\ \emph {et~al.}(2013)\citenamefont
  {Mendoza}, \citenamefont {Succi},\ and\ \citenamefont
  {Herrmann}}]{Mendoza2013}%
  \BibitemOpen
  \bibfield  {author} {\bibinfo {author} {\bibfnamefont {M.}~\bibnamefont
  {Mendoza}}, \bibinfo {author} {\bibfnamefont {S.}~\bibnamefont {Succi}}, \
  and\ \bibinfo {author} {\bibfnamefont {H.~J.}\ \bibnamefont {Herrmann}},\
  }\href {\doibase 10.1038/srep03106} {\bibfield  {journal} {\bibinfo
  {journal} {Scientific Reports}\ }\textbf {\bibinfo {volume} {3}},\ \bibinfo
  {pages} {3106} (\bibinfo {year} {2013})}\BibitemShut {NoStop}%
\bibitem [{\citenamefont {Debus}\ \emph {et~al.}(2017)\citenamefont {Debus},
  \citenamefont {Mendoza}, \citenamefont {Succi},\ and\ \citenamefont
  {Herrmann}}]{Debus2017}%
  \BibitemOpen
  \bibfield  {author} {\bibinfo {author} {\bibfnamefont {J.-D.}\ \bibnamefont
  {Debus}}, \bibinfo {author} {\bibfnamefont {M.}~\bibnamefont {Mendoza}},
  \bibinfo {author} {\bibfnamefont {S.}~\bibnamefont {Succi}}, \ and\ \bibinfo
  {author} {\bibfnamefont {H.~J.}\ \bibnamefont {Herrmann}},\ }\href {\doibase
  10.1038/srep42350} {\bibfield  {journal} {\bibinfo  {journal} {Scientific
  Reports}\ }\textbf {\bibinfo {volume} {7}},\ \bibinfo {pages} {42350}
  (\bibinfo {year} {2017})}\BibitemShut {NoStop}%
\bibitem [{\citenamefont {Bhatnagar}\ \emph {et~al.}(1954)\citenamefont
  {Bhatnagar}, \citenamefont {Gross},\ and\ \citenamefont
  {Krook}}]{Bhatnagar1954}%
  \BibitemOpen
  \bibfield  {author} {\bibinfo {author} {\bibfnamefont {P.~L.}\ \bibnamefont
  {Bhatnagar}}, \bibinfo {author} {\bibfnamefont {E.~P.}\ \bibnamefont
  {Gross}}, \ and\ \bibinfo {author} {\bibfnamefont {M.}~\bibnamefont
  {Krook}},\ }\href {\doibase 10.1103/PhysRev.94.511} {\bibfield  {journal}
  {\bibinfo  {journal} {Physical Review}\ }\textbf {\bibinfo {volume} {94}},\
  \bibinfo {pages} {511} (\bibinfo {year} {1954})}\BibitemShut {NoStop}%
\bibitem [{Note1()}]{Note1}%
  \BibitemOpen
  \bibinfo {note} {In lattice-Boltzmann models for fluid mechanics, this
  distribution function is proportional to the probability to find a molecule
  at position $\protect \mathaccentV {vec}17E{r}$ and time $t$ with velocity
  $\protect \mathaccentV {vec}17E{\xi }_i$, but it is just a system's variable
  in the general case}\BibitemShut {NoStop}%
\bibitem [{\citenamefont {Kr{\"{u}}ger}\ \emph {et~al.}(2017)\citenamefont
  {Kr{\"{u}}ger}, \citenamefont {Kusumaatmaja}, \citenamefont {Kuzmin},
  \citenamefont {Shardt}, \citenamefont {Silva},\ and\ \citenamefont
  {Viggen}}]{Kruger2017}%
  \BibitemOpen
  \bibfield  {author} {\bibinfo {author} {\bibfnamefont {T.}~\bibnamefont
  {Kr{\"{u}}ger}}, \bibinfo {author} {\bibfnamefont {H.}~\bibnamefont
  {Kusumaatmaja}}, \bibinfo {author} {\bibfnamefont {A.}~\bibnamefont
  {Kuzmin}}, \bibinfo {author} {\bibfnamefont {O.}~\bibnamefont {Shardt}},
  \bibinfo {author} {\bibfnamefont {G.}~\bibnamefont {Silva}}, \ and\ \bibinfo
  {author} {\bibfnamefont {E.~M.}\ \bibnamefont {Viggen}},\ }\href {\doibase
  10.1007/978-3-319-44649-3} {\emph {\bibinfo {title} {{The Lattice Boltzmann
  Method}}}},\ Graduate Texts in Physics\ (\bibinfo  {publisher} {Springer
  International Publishing},\ \bibinfo {address} {Cham},\ \bibinfo {year}
  {2017})\BibitemShut {NoStop}%
\bibitem [{\citenamefont {Lawden}(2002)}]{Lawden2002}%
  \BibitemOpen
  \bibfield  {author} {\bibinfo {author} {\bibfnamefont {D.~F.}\ \bibnamefont
  {Lawden}},\ }\href@noop {} {\emph {\bibinfo {title} {{Introduction to tensor
  calculus, relativity, and cosmology}}}}\ (\bibinfo  {publisher} {Dover
  Publications},\ \bibinfo {year} {2002})\ p.\ \bibinfo {pages}
  {205}\BibitemShut {NoStop}%
\bibitem [{\citenamefont {Thampi}\ \emph {et~al.}(2012)\citenamefont {Thampi},
  \citenamefont {Ansumali}, \citenamefont {Adhikari},\ and\ \citenamefont
  {Succi}}]{Thampi2012}%
  \BibitemOpen
  \bibfield  {author} {\bibinfo {author} {\bibfnamefont {S.~P.}\ \bibnamefont
  {Thampi}}, \bibinfo {author} {\bibfnamefont {S.}~\bibnamefont {Ansumali}},
  \bibinfo {author} {\bibfnamefont {R.}~\bibnamefont {Adhikari}}, \ and\
  \bibinfo {author} {\bibfnamefont {S.}~\bibnamefont {Succi}},\ }\href
  {\doibase 10.1016/j.jcp.2012.07.037} {\  (\bibinfo {year} {2012}),\
  10.1016/j.jcp.2012.07.037}\BibitemShut {NoStop}%
\bibitem [{\citenamefont {Rossing}\ \emph {et~al.}(2002)\citenamefont
  {Rossing}, \citenamefont {Wheeler},\ and\ \citenamefont
  {Moore}}]{Rossing2002}%
  \BibitemOpen
  \bibfield  {author} {\bibinfo {author} {\bibfnamefont {T.~D.}\ \bibnamefont
  {Rossing}}, \bibinfo {author} {\bibfnamefont {P.~P.~A.}\ \bibnamefont
  {Wheeler}}, \ and\ \bibinfo {author} {\bibfnamefont {F.~R.}\ \bibnamefont
  {Moore}},\ }\href@noop {} {\emph {\bibinfo {title} {{The science of
  sound}}}}\ (\bibinfo  {publisher} {Addison Wesley},\ \bibinfo {year} {2002})\
  p.\ \bibinfo {pages} {783}\BibitemShut {NoStop}%
\bibitem [{\citenamefont {Li}\ \emph {et~al.}(2016)\citenamefont {Li},
  \citenamefont {Zhou},\ and\ \citenamefont {Yan}}]{Li2016}%
  \BibitemOpen
  \bibfield  {author} {\bibinfo {author} {\bibfnamefont {Q.}~\bibnamefont
  {Li}}, \bibinfo {author} {\bibfnamefont {P.}~\bibnamefont {Zhou}}, \ and\
  \bibinfo {author} {\bibfnamefont {H.~J.}\ \bibnamefont {Yan}},\ }\href
  {\doibase 10.1103/PhysRevE.94.043313} {\bibfield  {journal} {\bibinfo
  {journal} {Physical Review E}\ }\textbf {\bibinfo {volume} {94}},\ \bibinfo
  {pages} {043313} (\bibinfo {year} {2016})}\BibitemShut {NoStop}%
\end{thebibliography}%

\section{Appendix}
\subsection{Chapman-Enskog Expansion for a forced lattice-Boltzmann}\label{ChapmannEnskogAppendix}
In this appendix we will show how the microdynamics of the lattice-Boltzmann equation, including a forcing term, can lead to a set of conservation laws with source terms like Eqs. \eqref{1Conservationcurved} and \eqref{2ConservationcurvedForce} for any lattice-Boltzmann scheme, in particular, the forces considered in this work. We will obtain Eq. \eqref{MomentumCurvedSystem} as the momentum correction needed to include such forcing terms in the lattice-Boltzmann scheme. The approach here described is based on the analysis made by Li, Zhou and Yan in 2016 \cite{Li2016}

Let us start by writing the lattice-Boltzmann equation with an additional forcing term discrete in the velocity space, $F_i$ 
\begin{equation}
\begin{split}
f_i\left(\vec{x}+\vec{\xi}_i\delta_{t},\,\vec{\xi}_i,\, t+\delta_{t}\right)-f_i\left(\vec{x},\,\vec{\xi}_i,\, t\right)=\\
-\frac{\delta_{t}}{\tau}\left[f_i\left(\vec{x},\,\vec{\xi}_i,\, t\right)-f_i^{eq}\left(\vec{x},\,\vec{\xi}_i,\, t\right)\right]+F_i\, ,\label{eq:Forcedevolution}
\end{split}
\end{equation}
now, let us perform the Chapman-Enskog expansion with this additional term. First, we expand the left-hand side of the Eq. \eqref{eq:Forcedevolution} in a Taylor series up to second order which gives
\begin{equation}
\begin{split}
\left[\frac{\partial}{\partial t} +\vec \xi_i\cdot\vec\nabla\right]f_i+\frac{1}{2}\left[\frac{\partial}{\partial t} +\vec \xi_i\cdot\vec\nabla\right]^2f_i+...\\
=\frac{1}{\tau}\left[f_i-f_i^{eq}\right]+F_i\, ,
\end{split}\label{TaylorForced}
\end{equation}
where $f_i=f_i\left(\vec{x},\,\vec{\xi}_i,\, t\right)$ and $\delta_t=1$. 
Second, we make a perturbative expansion in the small parameter $\epsilon$ of the distribution functions, the differential operators and the forcing term up to second order as follows $f_i=f_i^{(0)}+\epsilon f_i^{(1)}+\epsilon^2 f_i^{(2)}$, $\nabla=\epsilon\nabla_1$, $\frac{\partial}{\partial t}=\epsilon\frac{\partial}{\partial t_1} +\epsilon^2\frac{\partial}{\partial t_2}$ and $F_i=\epsilon F^{(1)}_i$. By replacing the perturbative expansions in Eq. \eqref{TaylorForced}, we get
\begin{equation}
\begin{split}
\left[\epsilon\frac{\partial}{\partial t_1} +\epsilon^2\frac{\partial}{\partial t_2} +\epsilon\vec \xi_i\cdot\vec\nabla_1\right]\left(f_i^{(0)}+\epsilon f_i^{(1)}+\epsilon^2 f_i^{(2)}\right)\\
+\frac{1}{2}\left[\epsilon\frac{\partial}{\partial t_1} +\epsilon^2\frac{\partial}{\partial t_2}+\epsilon\vec \xi_i\cdot\vec\nabla_1\right]^2\left(f_i^{(0)}+\epsilon f_i^{(1)}+\epsilon^2 f_i^{(2)}\right)+...\\
=\frac{1}{\tau}\left[\left(f_i^{(0)}+\epsilon f_i^{(1)}+\epsilon^2 f_i^{(2)}\right)-f_i^{eq}\right]+\epsilon F^{(1)}_i\, ,
\end{split}\label{TaylorForcedReplaced}
\end{equation}
equating order by order, we obtain for the zero-th first and second order, respectively
\begin{equation}
f_i^{(0)}=f_i^{(eq)}\quad,\label{ZeroOrderApp}
\end{equation}
\begin{equation}
\left[\frac{\partial}{\partial t_1} +\vec \xi_i\cdot\vec\nabla_1\right] f_i^{(0)}=-\frac{1}{\tau}f_i^{(1)}+F^{(1)}_i\quad,
\label{FirstOrderApp}
\end{equation}

\begin{equation}
\begin{split}
\frac{\partial}{\partial t_2}f_i^{(0)}+\left[\frac{\partial}{\partial t_1} +\vec \xi_i\cdot\vec\nabla_1\right] f_i^{(1)}\\
+\frac{1}{2}\left[\frac{\partial}{\partial t_1} +\vec \xi_i\cdot\vec\nabla_1\right]^2 f_i^{(0)}=\frac{1}{\tau}f_i^{(2)}\, ,
\end{split}
\label{SecondOrderApp}
\end{equation}
by using the Eq. \eqref{FirstOrderApp} we can write Eq.\eqref{SecondOrderApp} in the form
\begin{equation}
\begin{split}
\frac{\partial}{\partial t_2}f_i^{(0)}+\left[\frac{\partial}{\partial t_1} +\vec \xi_i\cdot\vec\nabla_1\right] f_i^{(1)}\\
+\frac{1}{2}\left[\frac{\partial}{\partial t_1} +\vec \xi_i\cdot\vec\nabla_1\right]\left(\frac{1}{\tau}f_i^{(1)}+F^{(1)}_i\right)=\frac{1}{\tau}f_i^{(2)}\, ,
\end{split}
\label{SecondApp}
\end{equation}
or, equivalently 
\begin{equation}
\begin{split}
\frac{\partial}{\partial t_2}f_i^{(0)}+\left(1-\frac{1}{2\tau}\right)\left[\frac{\partial}{\partial t_1} +\vec \xi_i\cdot\vec\nabla_1\right] f_i^{(1)}\\
+\frac{1}{2}\left[\frac{\partial}{\partial t_1} +\vec \xi_i\cdot\vec\nabla_1\right]F^{(1)}_i=\frac{1}{\tau}f_i^{(2)}\, .
\end{split}
\label{SecondOrderCompleteApp}
\end{equation}
If we sum Eqs. \eqref{FirstOrderApp} and \eqref{SecondOrderCompleteApp} over $i$ and use the definitions of the macroscopic variables (Eq. \eqref{MacroscopicQuantities}) and the Eq. \eqref{ZeroOrderApp}, we obtain
\begin{equation}\label{cons1app}
\frac{\partial}{\partial t_1} P + \vec{\nabla}_1 \cdot \vec{J}=0
\end{equation}
\begin{equation}\label{cons2app}
\frac{\partial}{\partial t_2} P + \vec{\nabla}_1 \cdot \frac{1}{2}\vec{\mathcal{F}}^{(1)}=0
\end{equation}
where we have assumed $\sum_i F_i=0$ and $\sum_i \vec{\xi}_i F_i=\vec{\mathcal{F}}$ as suggested by Guo \textit{et. al.}\cite{Guo2002}.
By adding the Eqs.\eqref{cons1app} and \eqref{cons2app} the first conservation law for the lattice-Boltzmann scheme reads
\begin{equation}\label{FirstConservationForcedApp}
\frac{\partial}{\partial t} P + \vec{\nabla}_1 \cdot \vec{J}'=0
\end{equation}
with
\begin{equation}\label{JShifted}
\vec{J}'=\vec{J}+\frac{1}{2}\vec{\mathcal{F}}
\end{equation}
Similarly, if we multiply by $\vec{\xi_i}$ before summing over $i$, we obtain the conservation laws of higher order
\begin{equation}
\frac{\partial}{\partial t_1} \vec{J} + \vec{\nabla}_1 \cdot \vec{\Pi^{(0)}}=\vec{\mathcal{F}}
\end{equation}
\begin{equation}
\begin{split}
\frac{\partial}{\partial t_1} \vec{J} + \left(1-\frac{1}{2\tau}\right)\vec{\nabla}_1 \cdot \vec{\Pi^{(1)}}\\
+\frac{1}{2}\frac{\partial}{\partial t_1}\vec{\mathcal{F}}^{(1)}+\frac{1}{2}\vec{\nabla}_1\left(\sum_i\vec{\xi}_i\vec{\xi}_iF^{(1)}_i\right)=0
\end{split}
\end{equation}
by summing these two equations and taking the choice $\tau=\frac{1}{2}$ for the wave equation, we obtain
\begin{equation}\label{SeconConservationForcedApp}
\frac{\partial}{\partial t}\vec{J}'+\nabla\cdot\Pi^{(0)}=\vec{\mathcal{F}}
\end{equation}
Again, the functional form of the forcing term in this case have to be chosen in order to fulfil $\sum_i\vec{\xi}_i\vec{\xi}_iF^{(1)}_i=0$. 
Eqs. \eqref{FirstConservationForcedApp} and \eqref{SeconConservationForcedApp} are exactly the conservation laws we use for our forced lattice-Boltzmann scheme (Eqs. 
\eqref{1Conservationcurved} and \eqref{2ConservationcurvedForce}) plus a rescaling factor of the macroscopic variables due to the geometry $\sqrt{g}$. Note, in addition, that the Eq.\eqref{JShifted} corresponds to Eq. \eqref{MomentumCurvedSystem}.

Finally, for the case of the LBM for the wave equation a simplification can be made on Eq. \eqref{eq:Forcedevolution} when we insert the value of $\vec{J}'$ into the equilibrium distribution function i.e. $f_i^{eq}\left(P,\vec{J}\right)\to f_i^{eq}\left(P,\vec{J}'\right)$. Let us write the equilibrium distribution function in the form
\begin{equation}
f_i^{eq}\left(P,\vec{J}\right)\equiv f_i^{eq}\left(P,\vec{J}'\right)-\left(f_i^{eq}\left(P,\vec{J}'\right)- f_i^{eq}\left(P,\vec{J}\right)\right)\, ,
\end{equation}
which allow us to rewrite the Eq. \eqref{eq:Forcedevolution} as
\begin{equation}
\begin{split}
f_i\left(\vec{x}+\vec{\xi}_i\delta_{t},\,\vec{\xi}_i,\, t+\delta_{t}\right)-f_i\left(\vec{x},\,\vec{\xi}_i,\, t\right)=\\
-\frac{\delta_{t}}{\tau}\left[f_i\left(\vec{x},\,\vec{\xi}_i,\, t\right)-f_i^{eq}\left(P, \vec{J}'\right)\right]+\tilde F_i\, ,\label{eq:ForcedevolutionRewrited}
\end{split}
\end{equation}
where 
\begin{equation}
\tilde F_i=F_i-\frac{1}{\tau}\left(f_i^{eq}\left(P,\vec{J}'\right)- f_i^{eq}\left(P,\vec{J}\right)\right)\, , 
\end{equation}
and the first three moments of this force are
\begin{equation}
\sum_i \tilde F_i=0
\end{equation}
\begin{equation}
\sum_i \vec{\xi}_i \tilde F_i=\left(1-\frac{1}{2\tau}\right)\vec{\mathcal{F}} = 0 \text{ for } \tau=\frac{1}{2}
\end{equation}
\begin{equation}
 \sum_i\vec{\xi}_i\vec{\xi}_i\tilde F_i=0
\end{equation}
Therefore, it wont have effect on the computed macroscopic fields and we can write:
\begin{equation}
\begin{split}
f_i\left(\vec{x}+\vec{\xi}_i\delta_{t},\,\vec{\xi}_i,\, t+\delta_{t}\right)-f_i\left(\vec{x},\,\vec{\xi}_i,\, t\right)=\\
-\frac{\delta_{t}}{\tau}\left[f_i\left(\vec{x},\,\vec{\xi}_i,\, t\right)-f_i^{eq}\left(P, \vec{J}'\right)\right]\, ,\
\end{split}
\end{equation}
which is the evolution equation we use in our model.

\subsection{Fields derivatives computed using a cell configuration\cite{Thampi2012}}
\label{AppendixA}

In some cases, it is necessary to compute the derivative of any scalar or vectorial field during a lattice-Boltzmann calculation, this can be done by implementing a discretization scheme like finite differences scheme, however, it can either reduce the accuracy order of the overall scheme or increase the computational time and model complexity. However, there is an useful alternative to compute the gradients of the fields that are involved in the lattice-Boltzmann simulation. 

The procedure to find these gradients is described here. Let us start from a Taylor expansion of the field in the direction of a given discrete velocity from the discrete set used in the lattice-Boltzmann calculations. 
\begin{equation}
\phi\left(\vec{x}+\delta_t\vec{\xi}_i\right)=\phi\left(\vec{x}\right)+\delta_t\vec{\xi}_i\nabla\phi\left(\vec{x}\right)+\delta_t^2\frac{\xi_i^\alpha\xi_i^\beta}{2}\nabla_\alpha\nabla_\beta\phi\left(\vec{x}\right)+...\, ,
\end{equation}
now, we multiply the whole expansion by $w_i\vec{\xi}_i$
\begin{equation}\begin{split}
w_i\vec{\xi}_i\phi\left(\vec{x}+\vec{\xi}_i\right)=&w_i\vec{\xi}_i\phi\left(\vec{x}\right)+w_i\vec{\xi}_i\vec{\xi}_i\cdot\vec\nabla\phi\left(\vec{x}\right)\\
&+w_i\frac{\vec\xi\vec\xi\vec\xi:\vec\nabla\vec\nabla}{2}\phi\left(\vec{x}\right)+...\, ,\end{split}
\end{equation}
if we sum over $i$, we can use the isotropy conditions \eqref{Weights} to cancel the terms of odd order in $\vec{\xi}_i$  
\begin{equation}
\sum_i w_i\vec{\xi}_i\phi\left(\vec{x}+\vec{\xi}_i\right)=\sum_i w_i\vec{\xi}_i\vec{\xi}_i\cdot\vec\nabla\phi\left(\vec{x}\right)+\mathcal{O}\left(\delta x^4\right)
\end{equation}
 and use 
\begin{equation}
\sum_i w_i\vec{\xi}_i\vec{\xi}_i=\sum_iw_i\xi^{\alpha}_i\xi^{\beta}_i=\delta^{\alpha\beta}c_s^2
\end{equation}
to write
\begin{equation}
\sum_i w_i\vec{\xi}_i\phi\left(\vec{x}+\vec{\xi}_i\right)=c_s^2\vec\nabla\phi\left(\vec{x}\right)+\mathcal{O}\left(\delta x^4\right)
\end{equation}
finally, if we divide by $c_s^2$ we obtain the required expression to compute the gradient of a scalar field by using the same discretization scheme implemented in the lattice-Boltzmann.
\begin{equation}
\vec\nabla\phi=\frac{1}{c_s^2}\sum_iw_i\vec\xi_i\phi\left(\vec{x}+\vec{\xi}_i\right)+\mathcal{O}\left(\delta x^2\right)
\end{equation}
It can also be generalized to compute the gradient of any tensor $A^{\alpha\beta}$
\begin{equation}
\partial_\alpha A^{\alpha\beta}=\frac{1}{c_s^2}\sum_iw_i\xi_i^\alpha A^{\alpha\beta}\left(x^\alpha+\xi_i^\alpha\right)+\mathcal{O}\left(\delta x^2\right)
\end{equation}
Note that the order of accuracy is the same of the overall lattice-Boltzmann scheme, even if the chosen discrete velocities set is only second order as in our case (D3Q7). 
\subsection{\label{TheoreticalPipe}Radial modes of the pipe.}
In order to define correctly the boundary condition for the rigid walls of the pipe and find a theoretical expression for the vibrational modes along the radial direction, lets first write the differential equation for $r$ obtained from the variables separation of the wave equation

\begin{equation}
\left(\frac{\partial^2}{\partial r^2}+\frac{1}{r}\frac{\partial}{\partial r}+\zeta^2-\frac{m^2}{r^2}\right)P_r=0
\end{equation}

whose solution is 

\begin{equation}
P_r=A_rJ_m\left(\zeta r\right)+B_rY_m\left(\zeta r\right)
\label{SolutionRadial}
\end{equation}

where $J_m\left(\zeta r\right)$ and $Y_m\left(\zeta r\right)$ are the Bessel functions of first and second kind respectively, the constants $A_r$ and $B_r$ can be obtained from the boundary conditions at $r=0$ and $r=r_{max}$. For $r=0$, since the pressure have to be finite and the Bessel Function of the second kind $Y_m(\zeta r)$ diverges at $r=0$, we have $B_r=0$. Now replacing the Eq. \ref{SolutionRadial} into the Eq. \ref{Boundary}, we find the condition

\begin{equation}
\zeta r\frac{\partial}{\partial \zeta r} J_m(\zeta r_{max})+J_m(\zeta r_{max})=0
\end{equation}

which can be numerically solved for $\zeta$ to find the characteristic radial frequencies of the pipe. In order to simplify the above equation and its solution we use the property 

\begin{equation}
\zeta r J'_m(\zeta r)=mJ_m(\zeta r)-\zeta r J_{m+1}(\zeta r)
\end{equation}

to find the function

\begin{equation}
f(\zeta r_{max})=(m+1)J_m(\zeta r_{max})-\zeta RJ_{m+1}(\zeta r_{max})
\label{FunctionZeros}
\end{equation}

whose zeros correspond to $\zeta_l r_{max}$, where $\zeta_l$ is the $l$-th radial characteristic frequency of the pipe.\\

For the simulations done in this work, we assume axial symmetry of the waves and therefore we choose $m=0$, the Figure \ref{Fig.BesselCondition}   shows the graphic of the Eq. \ref{FunctionZeros} as a function of $\zeta r_{max}$ for $m=0$

\begin{figure}
\centering
\includegraphics[width=0.4\textwidth]{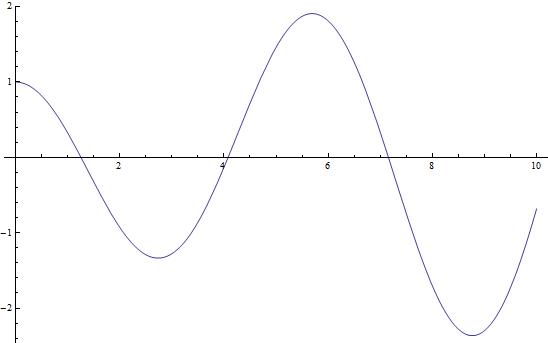} 
\caption{Graph of the function given by the Eq. \ref{FunctionZeros}}
\label{Fig.BesselCondition}
\end{figure}

The first three zeros of the function happen to be 

\begin{table}
\centering
\begin{tabular}{|c|c|}
\hline 
0rder & Value \\ 
\hline 
1 & 1,2556 \\ 
\hline 
2 & 4,0793 \\ 
\hline 
3 & 7,1390 \\ 
\hline
\end{tabular} 
\\
\caption{First three zeros of the Eq. \ref{FunctionZeros}}

\end{table}

Finally, the theoretical expression for the vibrational modes inside the pipe with $m=0$ considering both the radial and longitudinal vibrations is given by

\begin{equation}
\omega_{j,0,n}=c\left[\zeta_{l}^2+\left(\frac{(2n+1)\pi}{2L_z} \right )^2 \right ]^{1/2}\quad .
\label{TheoreticalOmega}
\end{equation}\\
\subsection{\label{AppHermite}Hermite expansion of the equilibrium distribution function}

Let us start by considering the conservation laws from the continuous Boltzmann equation. The statistical moments of the distributions are conserved under the collision process \cite{Kruger2017}; therefore, its value is the same whether they are computed from $f$ or from its equilibrium value $f^{eq}$,
\begin{equation}
\begin{split}
\int f d^3\xi&=\int f^{eq} d^3\xi=P\, ,\\
\int \vec{\xi} f d^3\xi&=\int \vec{\xi} f^{eq} d^3\xi=\vec{J}\, ,\\
\int \vec{\xi}\otimes\vec{\xi}fd^3\xi&=\int \vec{\xi}\otimes\vec{\xi}f^{eq}d^3\xi=\Pi^{\left(0\right)}\, .
\end{split}
\end{equation}
Our aim is to find a simpler form for this relations, but preserving the values of the macroscopic fields $P$, $\vec{J}$ and $\Pi^{\left(0\right)}$. In fluids, which are the macroscopic systems the Boltzmann equation is intended to model,  the equilibrium distribution is Maxwell-Boltzmann, which is a Gaussian distribution,
\begin{equation}
f^{eq}=\frac{\rho}{(\sqrt{2\pi RT})^D}\text{e}^{-(\vec{U}-\vec{\xi})^2/2RT}\, ,
\end{equation}
Where $\rho$ is the fluid density and $\vec{U}$ its velocity, $R=k_B/m$ is the ideal gas constant, with $k_B$ the Boltzmann constant and $m$, the mass of the particles; $T$, the temperature and $D$, the system dimension.
The clue is given by regarding that the continuous distribution function has the same functional form of the weight function for Hermite Polynomials,
\begin{equation}\label{WeightFunction}
w\left(x\right)=\frac{1}{\sqrt{2\pi}}\text{e}^{-x^2/2}\, ,
\end{equation}
which are defined by
\begin{equation}
\mathcal{H}^{(n)}\left(x\right)=\left(-1\right)^n\frac{1}{w\left(x\right)}\frac{d^n}{dx^n}w\left(x\right)\, ,
\end{equation}
and are orthogonal with the dot product 
\begin{equation}
f\cdot g = \int_{-\infty}^\infty w(x)f(x)g(x)dx\, .
\end{equation}

For the case of the wave equation, nevertheless, there is not a continuous distribution function that can be taken as a starting point to find its discrete version for LBM. So, we have to propose an appropriate continuous form to successfully retrieve the wave equation in the macroscopic limit. One can propose
\begin{equation}\label{continuousfeqapp}
f^{eq}\left(\vec{\xi}\right)=\frac{P}{(\sqrt{2\pi RT})^D}\text{e}^{-(\vec{J}-\vec{\xi})^2/2RT}\, ,
\end{equation}
\begin{equation}\label{continuousfeqGeneralapp}
f^{eq}\left(\vec{\xi}\right)=\frac{P}{(\sqrt{2\pi c_s^2})^D}\text{e}^{-[g_{\alpha\beta}\left(J^\alpha-\xi^\alpha\right)\left(J^\beta-\xi^\beta\right)]/2c_s^2}\, .
\end{equation}
for the Cartesian and generalized case respectively. Since the Hermite polynomials are orthogonal and form a basis, we can write any function $f(x)$ as a multidimensional Hermite polynomial series of the form 
\begin{equation}
f\left(x\right)=w\left(x\right)\sum_{n=0}^\infty\vec{a}^{\left(n\right)}\cdot\vec{\mathcal{H}}^{\left(n\right)}\left(x\right)\, ,
\end{equation}
where, the coefficients $\vec{a}^{\left(n\right)}$ can be obtained as
\begin{equation}
\vec{a}^{\left(n\right)}=\int f\left(x\right)\vec{\mathcal{H}}^{\left(n\right)}{\left(x\right)}dx\, .
\label{Coefficients}
\end{equation}
That is valid on each direction in velocity space, 
\begin{equation}
f^{eq}\left(\vec{\xi}\right)=w\left(\vec{\xi}\right)\sum_{n=0}^\infty\vec{a}^{\left(n\right)}\cdot\vec{\mathcal{H}}^{\left(n\right)}\left(\vec{\xi}\right)\, . 
\end{equation}
Note that one of the reasons for choosing Hermite polynomials is that the series coefficients directly correspond with the statistical moments of the distribution i. e. the system macroscopic quantities.
\begin{eqnarray}
&&a^{\left(0\right)eq}=\int f^{eq}d^d\xi=\rho=\int f d^d\xi\, ,\\
&&a^{\left(1\right)eq}=\int \xi^\alpha f^{eq} d^d\xi=J^\alpha=\int\xi^\alpha f d^d\xi\, .
\end{eqnarray}

Next, the expansion can be truncated up to certain order $N$ 
\begin{equation}\label{HermiteSeriesforF}
f^{eq}\left(\vec{\xi}\right)\approx w\left(\vec{\xi}\right)\sum_{n=0}^N\vec{a}^{\left(n\right)}\cdot\vec{\mathcal{H}}^{\left(n\right)}\left(\vec{\xi}\right)
\end{equation}
If we truncate the expansion up to second order we can write
\begin{equation}\label{SeriesExpansionFeqCartes}
f^{eq}\left(\vec{\xi}\right)=w\left(\vec{\xi}\right)\left[a^0 \mathcal{H}^0+a^1\mathcal{H}^1+a^2\mathcal{H}^2\right]\, .
\end{equation}
Another important feature of the Hermite polynomials is called Gauss-Hermite Cuadrature, which allow us to find exactly the value of an integral of a weighted polynomial of grade $n$, $P^{(n)}$, by considering a discrete sum over precise values $\xi_i$, which are actually the roots of the Hermite polynomial $\mathcal{H}^{(n)}$,
\begin{equation}
\int_{-\infty}^\infty w\left(\xi\right)P^{(n)}\left(\xi\right)d\xi=\sum_{i=1}^Nw_iP^{(n)}\left(\xi_i\right)\, ,
\label{GaussHermite}
\end{equation}
where $N$ must satisfy $n\geq\left(N+1\right)/2$. We can use the Eq. \eqref{GaussHermite} to easily compute the coefficients $\vec{a}^{\left(n\right)}$ (Eq. \eqref{Coefficients}),
\begin{equation}\label{defcoefcart}
\begin{split}
\vec{a}^{\left(n\right)}&=\int f^{eq}\left(\xi\right)\vec{\mathcal{H}}^{\left(n\right)}{\left(\xi\right)}d\xi\\
&=\int w\left(\xi\right)Q\left(\xi\right)\vec{\mathcal{H}}^{\left(n\right)}{\left(\xi\right)}d\xi=\sum_{i=1}^Nw_iQ\left(\xi_i\right)\vec{\mathcal{H}}^{\left(n\right)}\left(\xi_i\right)\, .
\end{split}
\end{equation}
The coefficients are
\begin{equation}\label{coefcartes}
\begin{array}{l l l}
a_0=P\, ,&\mathcal{H}_0=1\, ,\\
\vec{a}_1=\frac{\vec{J}}{c_s}\, ,&\mathcal{\vec{H}}_1=\frac{\vec{\xi}}{c_s}\, ,\\
a^{\alpha\beta}_2=\frac{P\left(c^2-c_s^2\right)\delta^{\alpha\beta}}{\sqrt{2}c_s^2}\, ,&\mathcal{H}^{\alpha\beta}_2=\frac{1}{\sqrt{2}}\left(\frac{\xi_i^\alpha\xi_i^\beta}{c_s^2}-\delta^{\alpha\beta}\right)\, .\\
\end{array}
\end{equation}
The corresponding equilibrium distribution function is 
\begin{equation}
f_{i\neq 0}^{eq}=w_i\left[P+\frac{\vec{\xi}_i\cdot\vec{J'}}{c_s^2}+\frac{P}{2c_s^4}\left(c^2-c_s^2\right)\left(\xi_i^2-3c_s^2\right)\right]\, .
\end{equation}
The expression for $f_0^{eq}$ can be obtained from 
\begin{equation}f_0^{eq}=P-\sum_{i=1}f_i^{eq}\, .
\end{equation} 
Summarizing, we have
\begin{equation}\label{AppEquilibriumHermiteCartesian}
f_i^{eq}=\begin{cases}
P-\left(\frac{5P}{2}-\frac{3c^2P}{2c_s^2}\right)\left(1-w_0\right)+\\
\frac{3c^2P}{2c_s^2}\left(c^2-c_s^2\right)&\text{if } i=0\\
\\
w_{i}\bigg[P+\frac{\vec{\xi}_i\cdot \vec{J}'}{c_s^2}\\
+\frac{P}{2c_s^4}\left(c^2-c_s^2\right)\left(\xi_i^2-3c_s^2\right)\bigg] & \mbox{otherwise}
\end{cases}\quad,
\end{equation}
\end{document}